\documentclass[twocolumn,tighten]{aastex701}
\usepackage{newtxtext,newtxmath}
\usepackage{amsmath}
\PassOptionsToPackage{breaklinks=true}{hyperref}
\def \deg      {$^{\circ}$}

\def\arcsec{$^{\prime\prime}$}

\received{June 25, 2026}
\revised{August 10, 2026}
\accepted{August 10, 2026}

\shorttitle{Extended radio emission in and around A\,520}
\shortauthors{Giacintucci et al.}

\begin{document}

\title{Extended radio emission in and around the merging cluster A\,520 as seen by MeerKAT}

\correspondingauthor{Simona Giacintucci}
\email{simona.giacintucci.civ@us.navy.mil}

\author{S. Giacintucci}
\affiliation{U.S. Naval Research Laboratory, 
4555 Overlook Avenue SW, Code 7213, 
Washington, DC 20375, USA}
\email{simona.giacintucci.civ@us.navy.mil}

\author{W. D. Cotton}
\affiliation{National Radio Astronomy Observatory, 520 Edgemont Road, Charlottesville, VA 22903, USA}
\affiliation{South African Radio Astronomy Observatory, 2 Fir Street, Black River Park, Observatory 7925, South Africa}
\email{x@y.z}

\author{M. Markevitch}
\affiliation{NASA/Goddard Space Flight Center,
Greenbelt, MD 20771, USA}
\email{maxim.markevitch@nasa.gov}

\author{T. Venturi}
\affiliation{INAF - Istituto di Radioastronomia,
via Gobetti 101, I-40129 Bologna, Italy}
\email{x@y.z}

\author{T. E. Clarke}
\affiliation{U.S. Naval Research Laboratory, 
4555 Overlook Avenue SW, Code 7213, 
Washington, DC 20375, USA}
\email{x@y.z}

\author{H. Bourdin}
\affiliation{Università di Roma Tor Vergata, Via della Ricerca Scientifica, I00133 Roma, Italy}
\affiliation{INFN, Sezione di Roma 2, Università di Roma Tor Vergata, Via della Ricerca Scientifica, 1, Roma, Italy}
\email{x@y.z}

\begin{abstract}
We present results from a deep MeerKAT 1.28 GHz observation of Abell 520, a galaxy cluster in the middle of a merger and a treasure trove of merger-generated phenomena. MeerKAT's combination of high angular resolution and sensitivity to extended radio emission lets us uncover unprecedented detail of the cluster's giant halo and puzzling features beyond the halo. We find that the main radio halo is sharply bounded along the direction of the NE-SW merger axis by the two X-ray shock fronts. In the perpendicular direction, the halo wings smoothly span over 2 Mpc, with the spectrum steepening in the outskirts. We detect faint radio emission upstream of the main (SW) bow shock; the emissivity jump across this shock can be used to constrain the physical mechanisms at work there. We discover a tail of radio emission extending past the NE shock to the gas-poor, high-entropy ``dark subcluster'' in the NE outskirts. The tail and subcluster have a higher radio/X-ray brightness ratio than the main halo's, suggesting efficient reacceleration of cosmic rays. We also discover two peripheral radio relics beyond the cluster virial radius, and a faint, $\sim2$ Mpc long, broad radio bridge connecting the dark subcluster to one of the relics along the direction perpendicular to the merger axis. The spectral slopes of the dark subcluster and the bridge are consistent with that of the main halo. This bridge and both relics may be caused by another giant merger-generated shock front at the virial radius.

\end{abstract}


\section{Introduction}
\label{sec:intro}

The dominant baryonic component in galaxy clusters is the intracluster medium (ICM) --- the hot ($kT\sim1-10$ keV), low-density plasma that emits X-rays via thermal bremsstrahlung. 
The ICM is permeated by magnetic fields and relativistic particles (or cosmic rays, hereafter CR), which generate radio synchrotron features \citep[see][for a review]{2019SSRv..215...16V}. Clusters observed during mergers often display cluster-wide (giant) radio halos and narrow peripheral radio relics. The likely mechanism behind the formation of halos is particle reacceleration due to merger-induced turbulence, while relics are believed to be produced by (re-)acceleration or adiabatic compression of fossil relativistic electrons by merger shocks \citep[see][for a review]{2014IJMPD..2330007B}. 
Relaxed clusters may exhibit diffuse radio emission in the form of smaller minihalos, which typically extend across the central cooling region \citep[e.g.,][]{2017ApJ...841...71G}. Minihalos are believed to arise from pre-existing, mildly relativistic seed electrons that are reaccelerated by turbulence associated with gas sloshing motions within the cluster core \citep[e.g.,][]{2013ApJ...762...78Z}. 
An alternative explanation for the origin of diffuse radio emission in clusters is offered by hadronic (or secondary) models, in which the radio-emitting electrons are produced locally through hadronic interactions between CR protons and the cluster thermal protons \citep[e.g.,][]{2004A&A...413...17P}. Hybrid models suggest that both primary (reaccelerated) and secondary CR electrons contribute to the diffuse radio emission \citep[e.g.,][]{2011MNRAS.412..817B,2017MNRAS.465.4800P,2025ApJ...978...62N}. 

Radio observations with the latest generation of highly sensitive, high-resolution radio interferometers (e.g., LOFAR\footnote{LOw Frequency ARray \citep{2013A&A...556A...2V}}, MeerKAT\footnote{\citep{2016mks..confE...1J,2018ApJ...856..180C}}, uGMRT\footnote{Upgraded Giant Metrewave Radio Telescope \citep{2017CSci..113..707G}}, JVLA\footnote{Jansky Very Large Array \citep{2011ApJ...739L...1P}}, and  ASKAP\footnote{Australia SKA Pathfinder \citep{2021PASA...38....9H}})  
are dramatically expanding our view of diffuse radio emission in clusters. These radio telescopes are revealing finer structures within diffuse radio sources, while simultaneously uncovering large-scale emission that was below the sensitivity limits of earlier instruments. This emission spans larger volumes and extends beyond halos and relics, with examples of faint emission bridges connecting merging clusters \citep[e.g.,][]{2019Sci...364..981G,2020MNRAS.499L..11B,2021ApJ...907...32B,2022A&A...660A..81V,2025A&A...694A.320H,2025A&A...695L..16S}, large-scale filamentary structures reaching out to the virial radius \citep[e.g.,][]{2022SciA....8.7623B,2026arXiv260714209B}, and diffuse emission that envelops the traditional radio halos \citep[e.g.,][]{2022Natur.609..911C,2025ApJ...992...88R,2026MNRAS.545f2123S}. These features indicate the presence of magnetic fields and relativistic particles out to very large cluster radii, in line with recent simulations suggesting that these components can be efficiently distributed throughout a significant portion of the cluster volume by gas bulk motions \citep{2021ApJ...914...73Z,2021A&A...653A..23V,2023A&A...669A..50V,2022MNRAS.510.4000F,2024Galax..12...19V} --- and allow us to use radio observations to diagnose the physical processes in other interesting ICM regimes.


\begin{table*}
\caption{MeerKAT observations of A\,520}
\begin{center}
\begin{tabular}{cccccccc}
\hline\noalign{\smallskip}
\hline\noalign{\smallskip}
Date & Block ID &  Track & On-source & Frequency & Bandwidth &   Primary & Secondary \\
     &                 &  (hour) & (hour) & (GHz) & (GHz) & calibrator & calibrator\\
\hline\noalign{\smallskip}
 2021-12-05 & 1638727749  & 8.8 & 7.5 & 1.28 & 0.86 & J\,0408--6545 & J\,0503+0203 \\
\hline{\smallskip} 
\end{tabular}
\end{center}
\label{tab:obs}
\tablecomments{Column 1: observation date. Column 2: capture block. Columns 3--4: total observing time (including calibration overheads) and time on target. Columns 5--6: central frequency and total bandwidth. Column 7: delay, bandpass and flux density calibrator. Column 8: phase calibrator.}
\end{table*}


In this paper, we present and analyze new, deep MeerKAT 1.28 GHz radio observations of Abell 520 (hereafter A\,520), a cluster at $z=0.203$ in the advanced stage of a merger in the plane of the sky. The cluster hosts a well-known Mpc-sized radio halo, which was first confirmed by \cite{2001A&A...376..803G} and further studied by \cite{2014A&A...561A..52V}, \cite{2018ApJ...856..162W} and \cite{2019A&A...622A..20H}, hereafter V14, W18 and H19. 
Besides the radio halo, A\,520 contains a number of unique features that make it an interesting laboratory for studying cluster physics. We provide an overview of these properties in Section~\ref{sec:a520}.

We adopt a $\Lambda$CDM cosmology with H$_0$=70 km s$^{-1}$ Mpc$^{-1}$, $\Omega_m=0.3$ and $\Omega_{\Lambda}=0.7$.  At the redshift of A\,520 ($z=0.203$), $1^{\prime\prime}$ corresponds to $3.34$ kpc. 
All errors are quoted at the $68\%$ confidence level. The radio spectral index $\alpha$ is defined according to $S_{\nu} \propto \nu^{-\alpha}$, where $S_{\nu}$ is the flux density at the frequency $\nu$.

\section{A\,520: a laboratory for cluster physics}
\label{sec:a520}

The merging galaxy cluster A\,520 is a collection of rare cluster phenomena. It contains one of the very few cluster bow shocks seen in the plane of the sky \citep[][hereafter M05 and W16]{2005ApJ...627..733M,2016ApJ...833...99W}, similar to the main shock front in the Bullet cluster
\citep{2002ApJ...567L..27M,2007PhR...443....1M}. 
The diffuse radio halo has a distinct, abrupt surface brightness edge that coincides with the bow shock (M05, W18), similar to a few other examples of radio halos with edges at merger shocks \cite[e.g.,][]{2010arXiv1010.3660M,2014MNRAS.440.2901S,2022ApJ...933..218B,2024A&A...690A.222B,2025ApJ...981..184S}. 

The {\em Chandra}\/ observations revealed a complex merger occurring roughly in the plane of the sky along the northeast–southwest direction that drives the bow shock, with additional evidence for a secondary merger component in the north–south direction (W16). A remnant cool core is identified near the X-ray peak, which has been disrupted and swept back by ram pressure during the merger, forming distinct structures referred to as the ``foot'' and ``knee'' in W16. In addition, a low X-ray surface brightness channel, 
extending $>$200 kpc along the north–south direction, has been detected by W16, likely associated with the secondary merger and possibly tracing regions of enhanced magnetic field
(i.e., a possible plasma depletion layer). 

The maps of the projected total mass distribution derived
using gravitational lensing revealed mass clumps in a chain along the main NE–SW merger axis indicated by the X-ray shock front \citep[][with an uncropped version of the same map presented in W16]{2007ApJ...668..806M, 2008PASJ...60..345O,2012ApJ...747...96J, 2014ApJ...783...78J,
2012ApJ...758..128C}.
 In particular, the map presented in W16 reveals two distinct subclusters outside the X-ray bright cluster region, about 0.8 and 1.2 Mpc northeast of the cluster center. They coincide with a low X-ray brightness tail in the {\em Chandra}\/ image. The outermost of those subclusters is particularly interesting: it is a fairly massive structure, with a total mass of $\sim 2.5–6\times10^{13}$ M$_{\odot}$ from the X-ray hydrostatic equilibrium estimate, broadly consistent with the weak-lensing mass estimate of $\sim 2\times10^{13}$ M$_{\odot}$. However, it has an anomalously low gas fraction of $f_{\rm gas}\lesssim0.03$ --- at least a factor two below typical gas fraction values at similar cluster radii (W16). 
This ``dark subcluster''%
\footnote{Note this is not the ``dark clump'' near the cluster center from \cite{2007ApJ...668..806M} and \cite{2012ApJ...747...96J}, which was not seen in the \cite{2012ApJ...758..128C} map.}
is most plausibly a stripped remnant of a subcluster that entered from the southwest 
but lost most of its original gas to ram-pressure stripping as it passed through the A\,520 core. The remaining dark matter clump then emerged on the northeast side, where it is now re-accreting and adiabatically compressing the surrounding high-entropy gas from the A\,520 outskirts, producing the faint X-ray enhancement seen in the X-ray images. This would explain the high specific entropy of the clump, which is similar to that of the cluster outskirts at that radius, rather than the lower entropy expected for a surviving cool core. W16 suggested that the second mass clump may be a similar object, although it is significantly affected by projection effects. 

We use the MeerKAT radio telescope, which combines high angular resolution, excellent aperture coverage and unprecedented sensitivity to low-surface-brightness radio emission, to investigate these X-ray structures in the radio band and look for new structures at larger linear scales.


\begin{table*}
\caption{Properties of the MeerKAT images}
\begin{center}
\begin{tabular}{ccccccccc}
\hline\noalign{\smallskip}
\hline\noalign{\smallskip}
\# & Frequency & Bandwidth & FWHM, p.a. & rms & Robust & $uv$ taper & Notes & Figure \\
 & (MHz)  &  (MHz) & ($^{\prime \prime} \times^{\prime \prime}$, $^{\circ}$) &  ($\mu$Jy beam$^{-1}$) &  & ($^{\prime\prime}$) &  & \\
 \hline\noalign{\smallskip}
1 & 1284 & 796 & $8.5 \times 6.3$, 156 & 3.1 & 0 & $-$ &  & \ref{fig:large}, \ref{fig:hr}{\em (a)}  \\
2 & 1284 & 796 & $8.5 \times 6.3$, 156 & 3.0 & 0 & $-$ &   & \ref{fig:hr}{\em (b)}, \ref{fig:xray}{\em (a)} \\
3 & 1284 & 796 & $23.3\times21.7$, 144 & 10.5 & 0 & 20  & & \ref{fig:hr}{\em (c)}, \ref{fig:corr}\\
4  & 1284 & 796 & $42.4\times40.8$, 138 & 23.3 & 0 & 40 & & \ref{fig:hr}{\em (d)} \\
5  & 1284 & 796 & $32.9\times31.2$, 140 & 14.5 & 0 & 30 & & \ref{fig:xray}{\em (b)}, \ref{fig:bow}{\em (c)}, \ref{fig:counter}{\em (c)}  \\
6  & 1284 & 796 & $10.0\times7.1$, 154 & 3.3 & 0.2 & $-$ &  & \ref{fig:bow}{\em (a)}, \ref{fig:counter}{\em (a)} \\
7 & 1035 & 298 & $34.6\times32.1$, 144 & 29.0 & 0 & 30 & $>0.1$ k$\lambda$ & $-$\\
8 & 1483 & 398 & $34.6\times32.1$, 144 & 15.9 & 0 & 30 & $>0.1$ k$\lambda$ & \ref{fig:spix}{\em (c,d)} \\
9 & 1035 & 298 & $13.1\times14.7$, 149 & 10.9 & 0 & 10 & $>0.1$ k$\lambda$ & $-$ \\
10 & 1483 & 398 & $14.7\times13.1$, 149 & 5.3 & 0 & 10 & $>0.1$ k$\lambda$ & \ref{fig:spix}{\em (a,b)} \\
\hline{\smallskip} 
\end{tabular}
\end{center}
\label{tab:images}
\tablecomments{Column 1: image number. Columns 2--3: post-flagging central frequency and bandwidth. Column 4: full width at half maximum (FWHM) and position angle (p.a.) of the radio beam. Column 5: rms noise level ($1\sigma$). Columns 6--7: Robust weighting and gaussian taper. Column 8: notes. Column 9: figure number. All images, except \#1, are source-subtracted. For the full-band images (\#1 to \#6), 99.5 MHz of the band, covering 1085–1184 MHz, was flagged due to contamination by RFI.}
\end{table*}

\begin{figure*}
\centering \epsscale{1.1}
\includegraphics[width=\hsize]{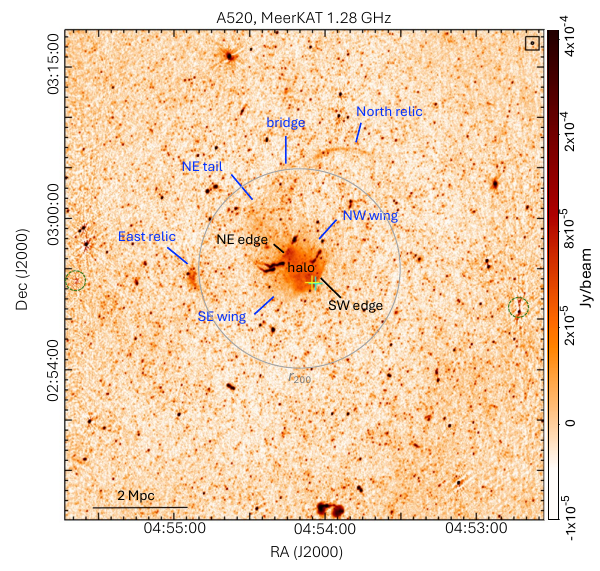}
\caption{MeerKAT 1.28 GHz image of A\,520 (image \# 1 in Tab.\ref{tab:images}) at the resolution of $8^{\prime\prime}.5 \times 6^{\prime\prime}.3$ (black, boxed ellipse). The noise is $1\sigma=3.1$ $\mu$Jy beam$^{-1}$. The region is $0^{\circ}.8$ in side, corresponding to a physical size of 9.6 Mpc. The cyan cross marks the position of the BCG, offset from the X-ray peak (yellow cross) by $20^{\prime\prime}$= 67 kpc. The gray circle is centered on the cluster X-ray centroid and has a radius of 2 Mpc, corresponding to $\approx r_{\rm 200}$. The cluster central region contains a previously-known giant radio halo, bounded by two sharp radio edges tracing X-ray shocks (W18, H19). Blue labels mark the new diffuse radio features unveiled by the MeerKAT observation. Green dashed circles mark the positions of two bright sources that were peeled from the data.} 
\label{fig:large}
\end{figure*}

\section{MeerKAT observations}\label{sec:obs}

We observed A\,520 with MeerKAT in L band (1.28 GHz) in December 2021 for a total of 7.5 hours on source (Program ID SCI-20210212-SG-01). The observations were conducted in full polarization and 4096$\times$208.984 kHz channel continuum mode with an 8-second integration time. The source J0408--6545 was used as the delay, bandpass, and flux density calibrator, while J0503+0203 served as phase calibrator. Table \ref{tab:obs} summarizes the observational details.

The data were calibrated using the $\tt OBIT$ package \citep{2008PASP..120..439C} following the procedure described in \cite{2020ApJ...888...61M}. This process included initial editing and flagging of data affected by radio frequency interference (RFI) and technical issues, followed by delay, bandpass, amplitude, and phase calibration. The flux density scale was set using the \cite{reynolds94} scale, with average residual amplitude calibration errors estimated to be within $5\%$. The data were then self-calibrated in {\tt OBIT} using the wide-band, wide-field imager MFImage\footnote{https://www.cv.nrao.edu/$\sim$bcotton/ObitDoc/MFImage.pdf}. The full band was sub-divided into eight frequency sub-bands\footnote{The third frequency sub-band (1085--1184 MHz) was entirely flagged due to RFI and excluded from imaging and subsequent analysis.}, each 99.5 MHz wide, which were imaged independently, using multiple facets to correct for sky curvature, and deconvolved jointly. We fully imaged an area of $1^{\circ}.2$ in radius around the phase center, adding smaller outlying facets for bright sources out to a radius of $1^{\circ}.8$. The self-calibration process included three phase-only iterations with a 30--second solution interval. To minimize image artifacts caused by two strong compact sources within the field of view, we implemented direction-dependent calibration by manually "peeling" these sources using MFImage\footnote{https://www.cv.nrao.edu/$\sim$bcotton/ObitDoc/ManualPeel.pdf}.  

After self-calibration, we converted the data set into a measurement set using CASA\footnote{Common Astronomy Software Applications, \citep{2022PASP..134k4501C}} and made our final images with WSClean, employing joint-channel and multi-scale deconvolution \citep{2014MNRAS.444..606O, 2017MNRAS.471..301O}. We subdivided the full bandwidth into 16 channels and applied a fourth-order polynomial fitting to account for wide--band spectral effects. 

Following \cite{2024A&A...690A.222B}, we extracted an inner region of the field of view by subtracting from the visibility data all sources above the $3\sigma$ level located outside a central squared region of 0.$^{\circ}5$ $\times$ 0.$^{\circ}5$ in size. We predicted their model visibility with WSClean\footnote{https://wsclean.readthedocs.io/en/latest/prediction.html} and then subtracted them from the visibility data set. This allowed us to limit the final imaging to a smaller region of interest centered on A\,520, while simultaneously reducing computational times. 

Our deepest image, obtained using a Briggs robustness weighting of 0 \citep{1995PhDT.......238B}, has a resolution of $8^{\prime\prime}.5\times6^{\prime\prime}.3$ and a noise level of $\sim 3$ $\mu$Jy beam$^{-1}$. This value exceeds the estimated confusion limit of $\sim 1.7$ $\mu$Jy beam$^{-1}$, indicating that the image 
is not yet confusion-limited and is dominated by thermal noise.

We also created images of the diffuse component at a lower angular resolution after subtracting the compact sources. These were first identified on a high-resolution image produced using only baselines longer than 400 $\lambda$ ($\sim 8^{\prime}.6$) to filter out the most extended emission, a Briggs robust parameter of $-1$, and the multi-scale option, with scales restricted to delta components and the synthesized beam \citep{2024A&A...690A.222B}. While this procedure helps to identify the compact sources, the sharp morphology of some of the cluster diffuse emission (such as the southwestern edge) may cause small-scale diffuse emission to be represented by model components. We therefore inspected the resulting model and excluded any components associated with the cluster emission from the source model prior to visibility subtraction. The remaining components were then used to subtract the compact sources from the visibility data. The resulting dataset was imaged both at high angular resolution and at lower resolution by down-weighting the longest baselines (using {\tt $-$taper-gaussian} in WSClean) to emphasize the diffuse emission on large scales. Because of the complexity involved in subtracting extended radio sources, we did not subtract the brightest extended radio galaxies in the cluster region. 

Image details are summarized in Tab.~\ref{tab:images}, where all images, except \#1, are source-subtracted. The properties of all diffuse structures (largest size, total flux density, average surface brightness and in--band spectral index) were measured using the source-subtracted images at low resolution (\#5, 7, 8 in Tab.~\ref{tab:images}) after correction for the MeerKAT primary beam attenuation using the Astronomical Image Processing System 
{\citep[AIPS;][]{2003ASSL..285..109G}}, following \cite{2020ApJ...888...61M}. 
Details of these measurements are provided later in \S~\ref{sec:flux}.

\begin{figure*}
\centering
\includegraphics[width=\hsize]{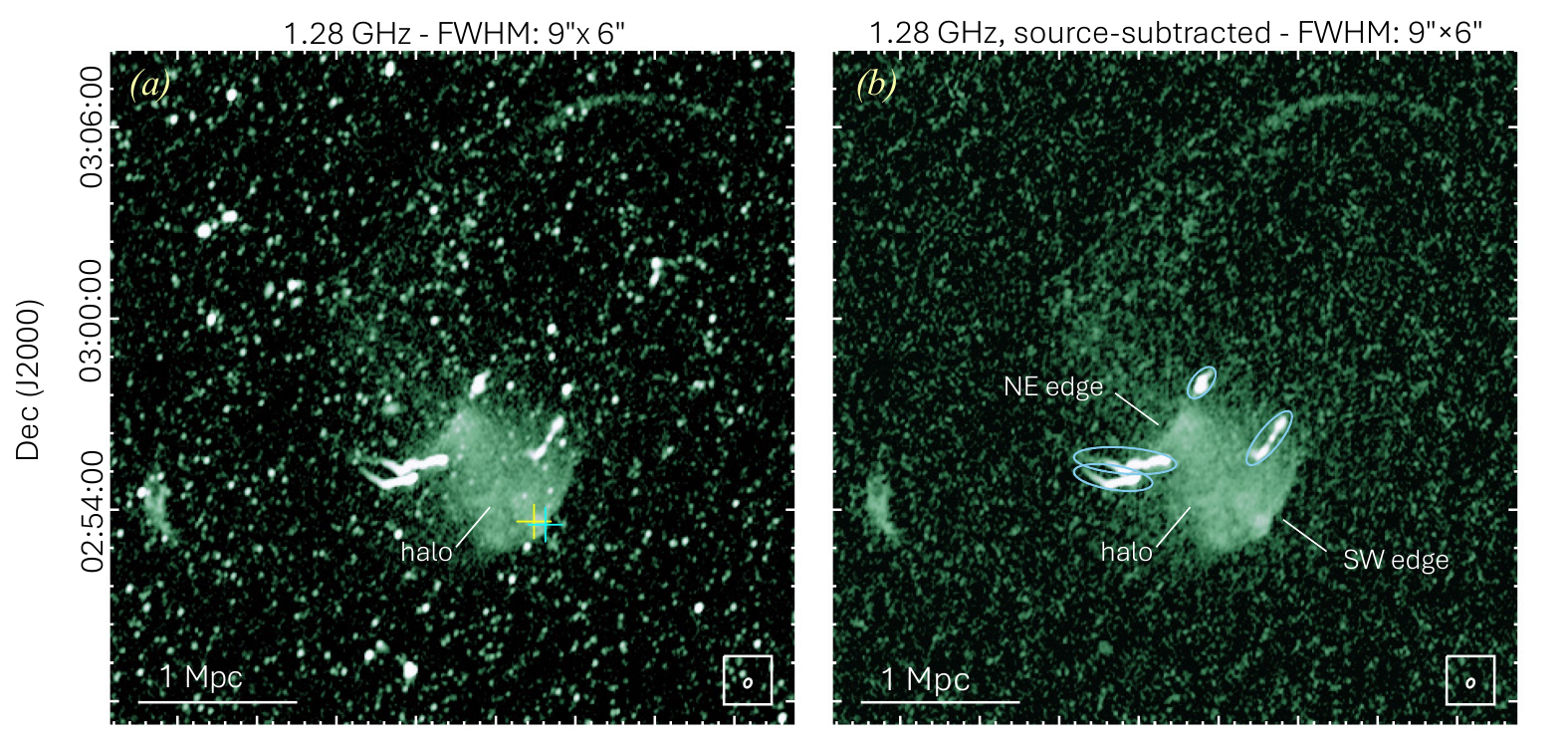}
\vspace{-0.3cm}
\includegraphics[width=\hsize]{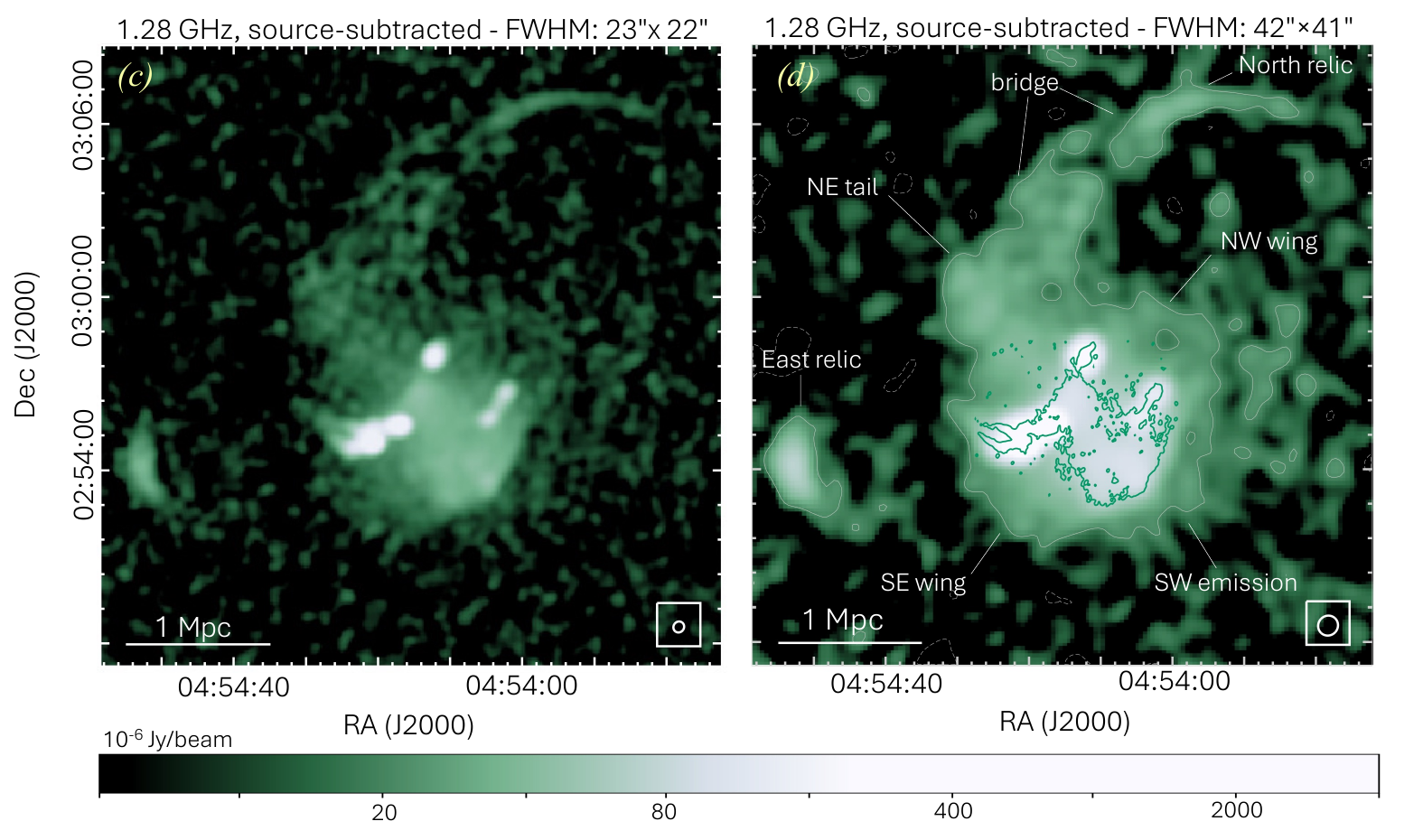}
\caption{MeerKAT 1.28 GHz images of A\,520. {\em (a)} Image at $8^{\prime\prime}.5\times6^{\prime\prime}.3$ resolution (\#1 in Tab.~\ref{tab:images}). The noise is $1\sigma=3.1$ $\mu$Jy beam$^{-1}$. The cyan cross marks the BCG location and the yellow cross marks the X-ray peak. {\em (b)} Same image after subtracting discrete sources (\#2 in Tab.~\ref{tab:images}); $1\sigma=3.0$ $\mu$Jy beam$^{-1}$.  Cyan ellipses mark four extended radio galaxies that were not subtracted. {\em (c)} Source-subtracted image at $23^{\prime\prime}.3\times21^{\prime\prime}.7$ resolution (\#3 in Tab.~\ref{tab:images}); $1\sigma=10.5$ $\mu$Jy beam$^{-1}$. {\em (d)} Source-subtracted image at $42^{\prime\prime}.4\times40^{\prime\prime}.8$ resolution (\#4 in Tab.~\ref{tab:images}); $1\sigma=23.3$ $\mu$Jy beam$^{-1}$. White contours are at $\pm2\sigma$ (solid/dotted). Green contours show the $5\sigma$ level from panel {\em (b)} and encompass the emission detected in previous VLA, GMRT and LOFAR images. In all panels, the white boxed ellipse indicates the radio beam.}
\label{fig:hr}
\end{figure*}

\section{MeerKAT radio images}\label{sec:images}

In Figure \ref{fig:large}, we present the MeerKAT full-band (0.9--1.7 GHz) image of A\,520, encompassing a region of $\sim10$ Mpc on a side. The position of the brightest cluster galaxy (BCG) is indicated by a cyan cross. The BCG is offset from the X-ray peak (yellow cross) by $\sim 20^{\prime\prime}$ = 67 kpc (W16). The gray circle is centered on the X-ray centroid and has a radius of 2 Mpc, roughly corresponding to $r_{\rm 200}$%
\footnote{$r_{200}$ is the radius within which the cluster mean total density is 200 times 
the critical density at the cluster redshift.}. 
The green dashed circles mark the position of the two bright sources that were peeled from the data.

The high angular resolution ($9^{\prime\prime}\times 6^{\prime\prime}$) and sensitivity ($\sim 3$ $\mu$Jy beam$^{-1}$ rms noise level) of this image allows us to map the previously-known, centrally-located radio halo with greater detail than earlier studies, while uncovering at the same time a wealth of newly-detected features (labeled in blue), extending to $r_{\rm 200}$ and beyond, that we describe in the next sections.

\subsection{The previously known radio halo}
\label{sec:halo}

Figure~\ref{fig:hr} presents MeerKAT 1.28~GHz images of A\,520 at multiple angular resolutions. At the highest resolution (panel {\em (a)}), the emission is dominated by the diffuse radio halo, which spans the central $\sim$1~Mpc region of the cluster, along with a large number of discrete radio galaxies. After subtraction of compact radio sources, the extended emission in the cluster center becomes more apparent (panels {\em (b,c,d)}). This includes the radio halo itself together with four bright, extended radio galaxies (indicated by cyan ellipses) that were not subtracted. 

In Figure~\ref{fig:xray}{\em (a)}, we overlay the radio contours from Fig.\ \ref{fig:hr}{\em (b)}\/ on the {\em Chandra}\/ X-ray image from W16. The radio halo is slightly elongated along the NE--SW merger axis. It shows an irregular and asymmetric surface brightness distribution, with a mild depression in the central region and the brightness increasing from the center toward the shock fronts (``bow shock'' and ``counter shock'' in Fig.\ \ref{fig:xray}{\em (a)}). There is a smaller-scale brightness peak on the inner side of the bow shock (see also Fig.~\ref{fig:hr}{\em (b)}), at the location where the X-ray image shows the ``foot'' and ``knee'' (W16; inset in Fig.~\ref{fig:xray}{\em (a)}) -- a chain of denser, cooler clumps of gas apparently remaining from the disrupted cool core of the infalling subcluster. Their association suggests that the peak of the radio emission there may be the remnant of a diffuse minihalo that inhabited the cool core prior to its disruption by the merger (W18). 

The radio halo emission is bounded by two distinct radio surface brightness edges (SW edge and NE edge), which appear remarkably sharp in the high-resolution MeerKAT images. These radio edges coincide with the X-ray bow shock and counter shock (Fig.~\ref{fig:xray}{\em (a)}; see also W18, H19). MeerKAT reveals faint diffuse emission beyond both these radio edges, which we discuss below.

\subsection{New diffuse emission}
\label{sec:new}

The high-resolution images in Figures \ref{fig:large} and \ref{fig:hr}{\em (a,b)}\/ hint at the presence of additional, fainter diffuse emission on scales larger than the central radio halo, which becomes apparent in the lower-resolution images in Fig.~\ref{fig:hr}{\em (c,d)} with increased sensitivity to the diffuse emission. The radio emission extends far beyond the inner halo region approximately demarcated by the green contour in panel {\em (d)}, which corresponds to the $5\sigma$ surface brightness level in the high-resolution MeerKAT image in panel {\em (b)}\/ and encompasses the $\sim 1$ Mpc halo previously seen by VLA, GMRT and LOFAR (V14, W18, H19). The emission can be traced to extremely low surface brightness levels ($\sim0.03-0.07$ $\mu$Jy\,arcsec$^{-2}$ --- an order of magnitude below the average surface brightness of the central giant halo, $\sim 0.35$ $\mu$Jy\,arcsec$^{-2}$), far below the sensitivity of the previous radio observations. It reveals a cluster significantly more complex than previously known.

\subsubsection{NW and SE wings}
\label{sec:extensions}

\begin{figure*}
\centering
\includegraphics[width=\hsize]{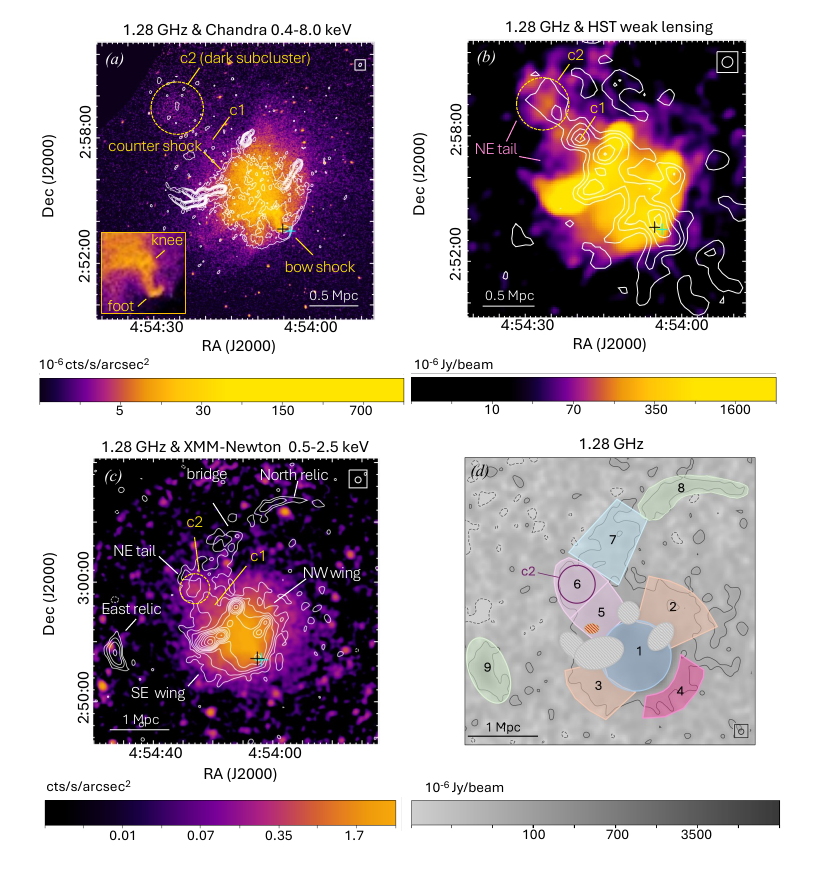}
\caption{{\em (a)} {\em Chandra} 0.8--4.0 keV image, binned to 1$^{\prime\prime}$ pixel and smoothed with a Gaussian ($\sigma = 1^{\prime\prime}$). Contours show the diffuse radio emission from  Fig.~\ref{fig:hr}{\em (b)}, starting at $+5\sigma$ and doubling thereafter. No $-5\sigma$ levels are present. {\em (b)} 1.28 GHz image at $32^{\prime\prime}.9\times31^{\prime\prime}.2$ resolution ($1\sigma=14.5$ $\mu$Jy beam$^{-1}$; \#5 in Tab.~\ref{tab:images}). Contours show the weak-lensing mass distribution from W16. {\em (c)} {\em XMM-Newton} 0.5–2.5 keV image, binned to 1$^{\prime\prime}$.7 pixel and smoothed with a Gaussian ($\sigma = 8^{\prime\prime}$.5). Radio contours (from {\em (b)}) double from $+3\sigma$. Dashed contours show the  $-3\sigma$ level. {\em (d)} Regions used to measure the properties of the diffuse components (Tab.~\ref{tab:diffuse}), overlaid on the same image as in panel {\em (b)}. Contours are at $\pm2\sigma$ (solid/dotted). Unsubtracted/partially subtracted extended sources are masked (gray/red dashed ellipses). In all panels, the boxed ellipse indicates the radio beam. The cyan and black crosses mark the BCG and X-ray peak location, respectively. }
\label{fig:xray}
\end{figure*}

The central halo extends smoothly, without sharp boundaries or edges, in the northwest and southeast directions, perpendicular to the merger axis (and the direction of the shocks). These NW-SE wings span at least $d\sim$2 Mpc --- double the size of the previous halo detection (Fig.\ \ref{fig:hr}{\em (d)}). In Fig.~\ref{fig:xray}{\em (c)}, we overlay the MeerKAT low-resolution image (with a beam size intermediate between panels {\em (c)}\/ and {\em (d)}\/ of Fig.\ \ref{fig:hr}) on the large-scale X-ray image from {\em XMM-Newton}. The NW and SE wings roughly follow the elongation of the thermal X-ray emission in the same direction.

\begin{figure}
\centering
\includegraphics[width=\hsize]{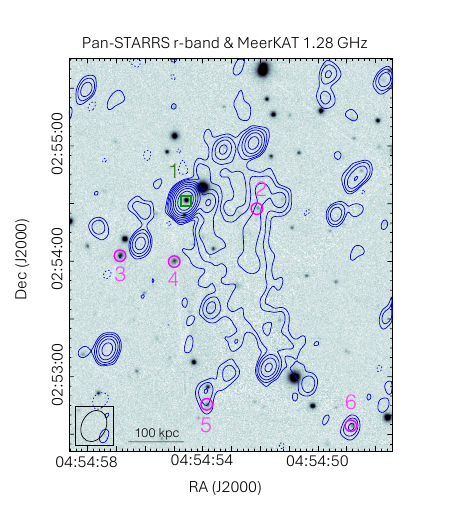}
\caption{Panoramic Survey Telescope \& Rapid Response System (Pan-STARRS; \cite{2016AAS...22732407C}) r-band image of the region occupied by the East relic. 
Contours show the MeerKAT high-resolution image at 1.28 GHz, presented in Fig.~\ref{fig:hr}{\em (a)}, starting at $+3\sigma$ and increasing by factors of two. Dashed contours indicate the  $-3\sigma$ level. The numbered magenta circles mark galaxies with redshift information
from \cite{2008A&A...491..379G} and NED, with redshifts of 0.232 (object marked as 2), 0.231 (object 3), 0.207 (object 4), 0.202 (object 5) and 0.206 (object 6). The object marked as 1 (green box) is a candidate quasar (Gaia DR2 3233066284519369728) for which no redshift is currently available.}
\label{fig:erelic}
\end{figure}

\subsubsection{The radio relics}
\label{sec:relics}

MeerKAT reveals two relic-like structures, located beyond the virial radius of A\,520 (Figs.\ \ref{fig:large} and ~\ref{fig:hr}). The North relic is found $\sim 2.4$ Mpc north of the X-ray centroid. It has a thin ($\lesssim 200$ kpc), arc-shaped morphology and spans a total projected length of $\sim 1.5$ Mpc, which falls in the range of sizes commonly reported for radio relics \citep[$\sim$0.5-2 Mpc; e.g.,][]{2019SSRv..215...16V}. 
A second, smaller relic (East relic in Figs.~\ref{fig:large},  \ref{fig:hr} and \ref{fig:xray}{\em (c)}) is found $\sim 2.2$ Mpc east of the X-ray centroid. It is less extended, with a size of $\sim$300 kpc $\times$ 800 kpc. As shown in Fig.~\ref{fig:erelic}, the diffuse radio emission shows no obvious optical counterpart or connection to a single host galaxy that would suggest a remnant radio galaxy. Several galaxies in the vicinity have available redshift measurements from 
\cite{2008A&A...491..379G} and the NASA/IPAC Extragalactic Database (NED) consistent with that of A\,520 (magenta circles), but they are spatially distinct from the diffuse radio emission. The bright compact radio source associated with object 1 (green box) is a candidate quasar identified in Gaia Data Release 2 (Gaia DR2 3233066284519369728; \cite{2018A&A...616A...1G}). These properties, together with its curved morphology, support its interpretation as a radio relic rather than a remnant radio galaxy.

Both relics are offset from the merger axis on the opposite sides of it. A comparison with the {\em XMM-Newton} X-ray image (Fig.~\ref{fig:xray}{\em (c)}) shows that both relics lie beyond the extent of the detected thermal X-ray emission from the cluster, at least with the current depth of the available X-ray observations.

\subsubsection{The NE tail and the bridge}
\label{sec:ne_tail}

A faint tail extends from the NE edge of the radio halo, marked by the counter shock, to the northeast along the main NE-SW merger axis of the cluster (NE tail in Fig.~\ref{fig:hr}). Its full extent (1.5 Mpc from the cluster center, 1.2 Mpc from the counter shock) and morphology are revealed in the lower-resolution images in
panels {\em (c,d)}, where the bulk of the diffuse emission, mostly resolved out at higher resolution, is detected at $\gtrsim 5\sigma$ significance. The NE radio brightness edge suggests that this tail is a distinct emission component rather than an extension of the halo. At its northern end, the emission bends abruptly by $\sim 90^{\circ}$ into a fainter bridge that connects, in projection, the NE tail and the near end of the North relic (Fig.~\ref{fig:hr}{\em (d)}). 

A comparison with the weak-lensing mass distribution map from W16 (Fig.~\ref{fig:xray}{\em (b)}) and {\em XMM-Newton} X-ray image (Fig.~\ref{fig:xray}{\em (c)}) shows that the NE tail is co-spatial with the mass clumps c1 and c2 identified by W16 and located at $\sim$0.8 Mpc and $\sim$1.2 Mpc from the cluster center along the merger axis. The NE tail is composed of two radio enhancements, each associated with one of these mass clumps. The outermost clump, c2, is the dark subcluster described in \S~\ref{sec:a520}. The spatial coincidence of the NE tail with these mass clumps, combined with its elongation along the merger axis, suggests a physical association. The detection of diffuse radio emission at the dark subcluster is particularly intriguing, as such massive, gas-stripped structures are extremely rare. We discuss the possible nature of the radio emission in \S~\ref{sec:disc_dark}.

The bridge between the NE tail and the North relic is $\sim$0.6 Mpc broad and 1 Mpc long (Fig.~\ref{fig:hr}{\em (d)}). 
It is one of the faintest components detected in the MeerKAT images (0.035 $\mu$Jy arcsec$^{-2}$).
It is unclear whether this bridge represents a physical connection or is a chance line-of-sight alignment, although its alignment with the relic is suggestive. We discuss its possible nature in \S~\ref{sec:disc_dark}.


\begin{table*}[tb]
\caption{Properties of the diffuse radio structures in A\,520.}
\begin{center}
\begin{tabular}{cccccc}
\hline\noalign{\smallskip}
\hline\noalign{\smallskip}
Region & Component & Size & $S_{\rm 1284 \, MHz}$ & $SB_{\rm average}$ & $\alpha$  \\
  &  & (Mpc$\times$Mpc)  &  (mJy) & ($\mu$Jy arcsec$^{-2}$) &  (1035--1483 MHz) \\ 
  \hline\noalign{\smallskip}
 1 & halo & $0.9\times1.0$ & $19.0\pm0.1(1.9)$ & 0.35 & $1.32\pm0.03(0.40)$    \\
 2 & NW wing & $0.5\times1.0$ & $1.47\pm0.10(0.18)$ & 0.027 & $1.6\pm0.4(0.6)$  \\
 3 & SE wing & $0.5\times0.9$&  $1.48\pm0.08(0.17)$ & 0.042 & $2.1\pm0.3(0.5)$  \\  
 4 & SW pre-shock & $0.3\times1.0$ & $0.78\pm0.08(0.11)$  &  0.025 & $2.2\pm0.6(0.8)$   \\
 5 & NE tail (c1)&  $0.5\times0.6$ & $1.83\pm0.07(0.20)$ & 0.070 & $1.3\pm0.2(0.5)$   \\
 6 & NE tail (c2) & $0.6\times0.6$ &  $1.93\pm0.08(0.21)$ & 0.056 & $1.5\pm0.2(0.5)$  \\
 7 & Bridge & $0.6\times1.0$& $2.18\pm0.11(0.24)$ & 0.035 &$ 1.4\pm0.3(0.5)$ \\
8 & North relic & $0.2\times1.5$&  $1.35\pm0.08(0.16)$ & 0.040 & $ 1.4\pm0.3(0.5)$ \\ 
9 & East relic & $0.3\times0.8$& $2.26\pm0.08(0.24)$ & 0.066 & $1.5\pm0.2(0.4)$ \\
\hline{\smallskip} 
\end{tabular}
\end{center}
\label{tab:diffuse}
\tablecomments{Column 1: region number from Fig.~\ref{fig:xray}{\em d}. Column 2: corresponding radio component. Column 3: largest linear extent. Column 4: flux density at 1284 MHz. Column 5: Average surface brightness. Column 6: in-band (1035--1483 MHz) spectral index. All measurements were done using the low-resolution images \#5, 7 and 8 in Tab.~\ref{tab:images} within the $2\sigma$ contour. Reported uncertainties are estimated from the image rms noise; values in parenthesis include also the combined systematic uncertainties from absolute flux density scale calibration and discrete sources subtraction. }
\end{table*}


\subsubsection{SW pre-shock emission}
\label{sec:preshock}

Figure~\ref{fig:hr}{\em (d)} reveals extremely faint ($\sim$0.03 $\mu$Jy\,arcsec$^2$) diffuse emission extending beyond the SW edge of the radio halo (SW emission). It is detected at the $\sim 2\sigma$ level out to $\sim$ 300 kpc from the halo edge, which is also the X-ray bow shock, and lies in the undisturbed, cooler gas upstream of the shock front. We refer to it as the SW pre-shock emission. In \S~\ref{sec:analysis}, we provide further support for this detection through radial surface brightness profiles extracted across the bow shock region, and discuss this detection in \S~\ref{sec:disc_preshock}.

\section{Flux density and spectral index measurements}
\label{sec:flux}

We measured the radio properties of the radio halo and newly-revealed diffuse structures in A\,520 using the MeerKAT source-subtracted image at low resolution (\#5 in Tab.~\ref{tab:images}), after correction for the primary beam attenuation (\S~\ref{sec:obs}). The regions used for these measurements are shown in Fig.~\ref{fig:xray}{\em (d)} (for the SW pre-shock region, we excluded the region immediately adjacent to the bow shock to avoid contamination from the halo). Table~\ref{tab:diffuse} summarizes the largest size, flux density and average surface brightness at the central frequency of 1284 MHz, and the in-band spectral index of all components. These latter were computed from flux density measurements obtained from a pair of images centered at 1035 MHz and 1483 MHz (\#7 and \#8 in Tab.~\ref{tab:images}) produced with matching beams of $34^{\prime\prime}.6\times32^{\prime\prime}.1$. We also imposed a minimum baseline cut of $0.1$ k$\lambda$ (corresponding to a largest recoverable scale of $\sim$4 Mpc at the cluster redshift) to ensure sensitivity to the same range of spatial scales. Both images were corrected for the primary beam attenuation prior to flux density integration. The errors reported in Table~\ref{tab:diffuse} were estimated from the image rms noise. The values in parentheses include also the combined systematic uncertainties from the absolute flux density scale calibration and discrete source subtraction, assumed to be 10\% \citep[see, e.g.,][]{2025ApJ...992...88R}.

The average surface brightness of all newly-uncovered components ranges from $\sim$0.03 to $\sim$0.07 $\mu$Jy arcsec$^{-2}$, which is roughly one order of magnitude below the average surface brightness of the halo (0.35 $\mu$Jy arcsec$^{-2}$), consistent with their non-detection in previous, less sensitive radio observations of A\,520. The total flux density of the radio halo at 1284 MHz is $19.0\pm1.9$ mJy, corresponding to a $k$-corrected radio power of $(2.1\pm0.2)\times10^{24}$ W Hz$^{-1}$ at 1.4 GHz using the in-band spectral index in Tab.~\ref{tab:diffuse}. This is consistent within errors with previous measurements by  \cite{2013ApJ...777..141C}, W18 and H19 using the VLA data. If we consider the NW and SE wings as extensions of the radio halo at larger radii, as suggested by their morphology and lack of sharp boundaries with the central halo region (\S~\ref{sec:extensions}), the total flux density of the whole halo+wing system would be 23.2 mJy (corresponding to $\sim $2.7$\times10^{24}$ W Hz$^{-1}$ at 1.4 GHz), which represents an increase of $\sim 22\%$ over the flux density of the inner halo alone. 

For the halo, H19 derived an integrated spectral index of $\alpha = 1.04 \pm 0.05$ based on a power-law fit to data at 145 MHz, 323 MHz and 1.5 GHz within the $3\sigma$ contour of their images. Our in-band value of $\alpha = 1.32 \pm 0.03$ ($\pm0.40$, including systematic uncertainties) for the halo is derived for a slightly larger circular region of $r=0.5$ Mpc (Fig.~\ref{fig:xray}{\em (d)}) and over a much narrower frequency range (1035--1483 MHz). The two values are still consistent within the larger uncertainties associated with our measurement when systematic errors are included. Spectral indices derived over a narrow frequency range represent the local spectral shape and can thus be different from the broad-band slope, for example, because of the spectral curvature.

The in-band spectral indices of the new components (Tab.~\ref{tab:diffuse}) are roughly consistent within their uncertainties with that of the radio halo, except for the SE wing and SW pre-shock regions that exhibit a steeper slope of $\alpha\sim2$. The spectral indices of all components in Tab.~\ref{tab:diffuse} are typical for diffuse cluster radio sources \citep[e.g.,][]{2019SSRv..215...16V}. However, we caution that our in-band measurements have large uncertainties because of the limited frequency coverage as well as systematic uncertainties from compact source subtraction affecting the lower surface brightness components. Wider frequency coverage will therefore be needed for a more robust spectral characterization.

\begin{figure*}
\centering
\includegraphics[width=\hsize]{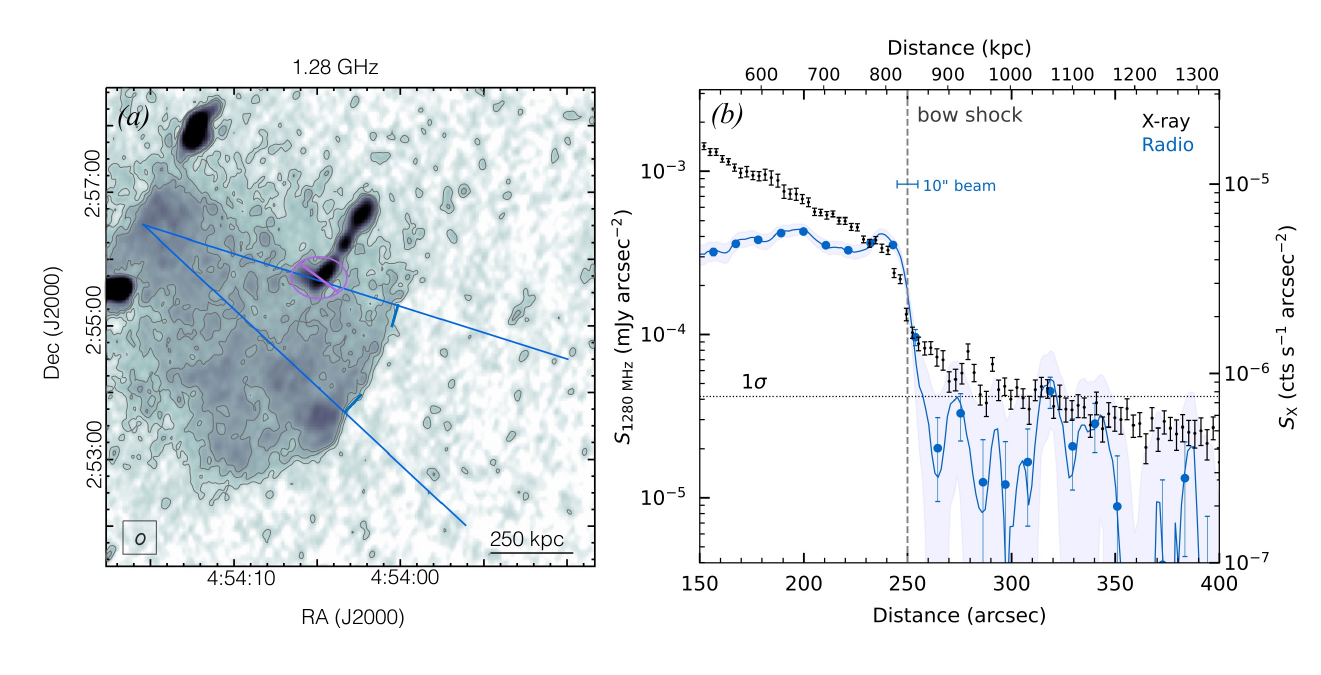}
\includegraphics[width=\hsize]{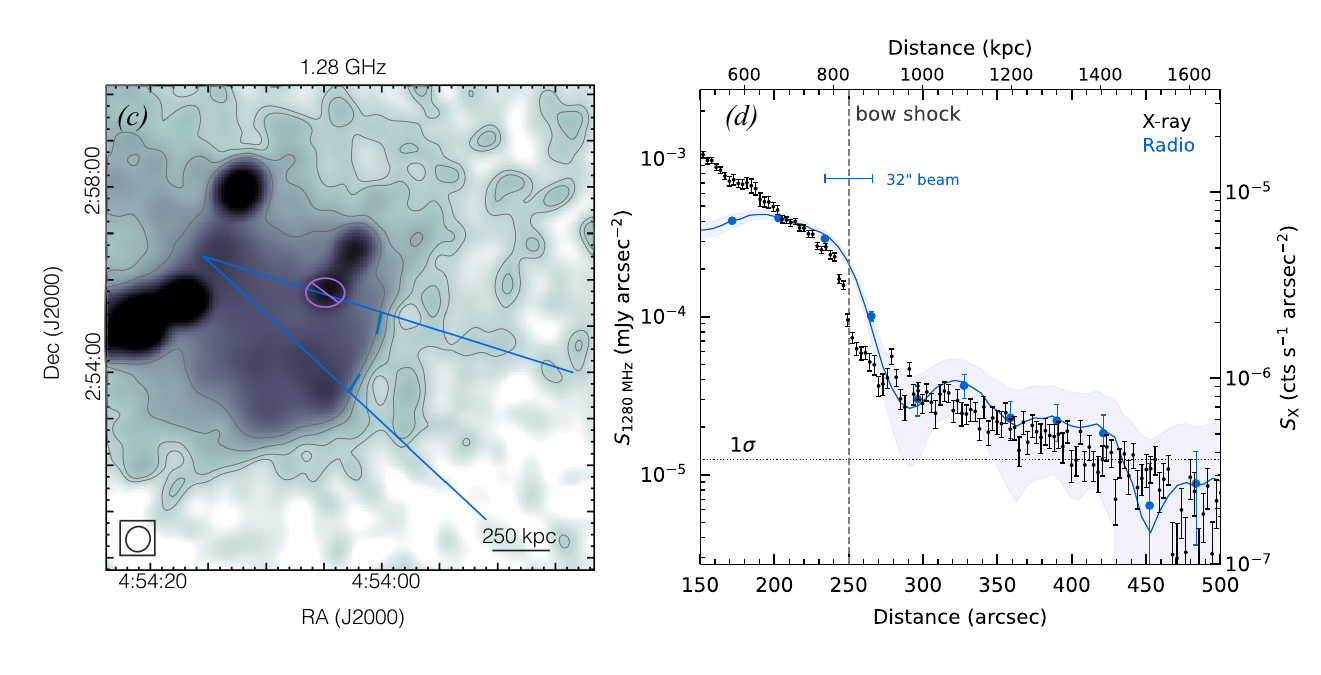}
\caption{Panels {\em (a,c)}: 1.28 GHz images of the southwest region of the radio halo at $10^{\prime\prime}\times7^{\prime\prime}$ 
and  $33^{\prime\prime}\times31^{\prime\prime}$ resolution (boxed ellipses; $1\sigma=3.3$ $\mu$Jy beam$^{-1}$ and $14.5$ $\mu$Jy beam$^{-1}$; \#6 and 5 in Tab.~\ref{tab:images}). Contours are at 3, 4, and 5$\sigma$ in panel {\em (a)} and 2, 3, and 5$\sigma$ in panel {\em (c)}. Blue lines indicate the sector used to extract the profiles in panels {\em (b)} and {\em (d)}; short ticks show the position of the X-ray shock front. Contaminating emission from an unsubtracted radio galaxy is masked (crossed ellipse). Panels {\em (b,d)}: Radio (blue) and X-ray (black) surface brightness profiles. The solid blue line and shaded region (1$\sigma$ uncertainties) are derived from radial bins of $1^{\prime\prime}.5$ in {\em (b)} and $6^{\prime\prime}$ in {\em (d)}. Blue points, with $1\sigma$ error bars, correspond to bins comparable to the beam size ($10^{\prime\prime}$ and $32^{\prime\prime}$, respectively). The horizontal dotted line indicates the radio 1$\sigma$ level, and the vertical dashed line marks the bow shock position (Fig.~\ref{fig:xray}{\em (a)}). 
} 
\label{fig:bow}
\end{figure*}

\begin{figure*}
\centering
\includegraphics[width=\hsize]{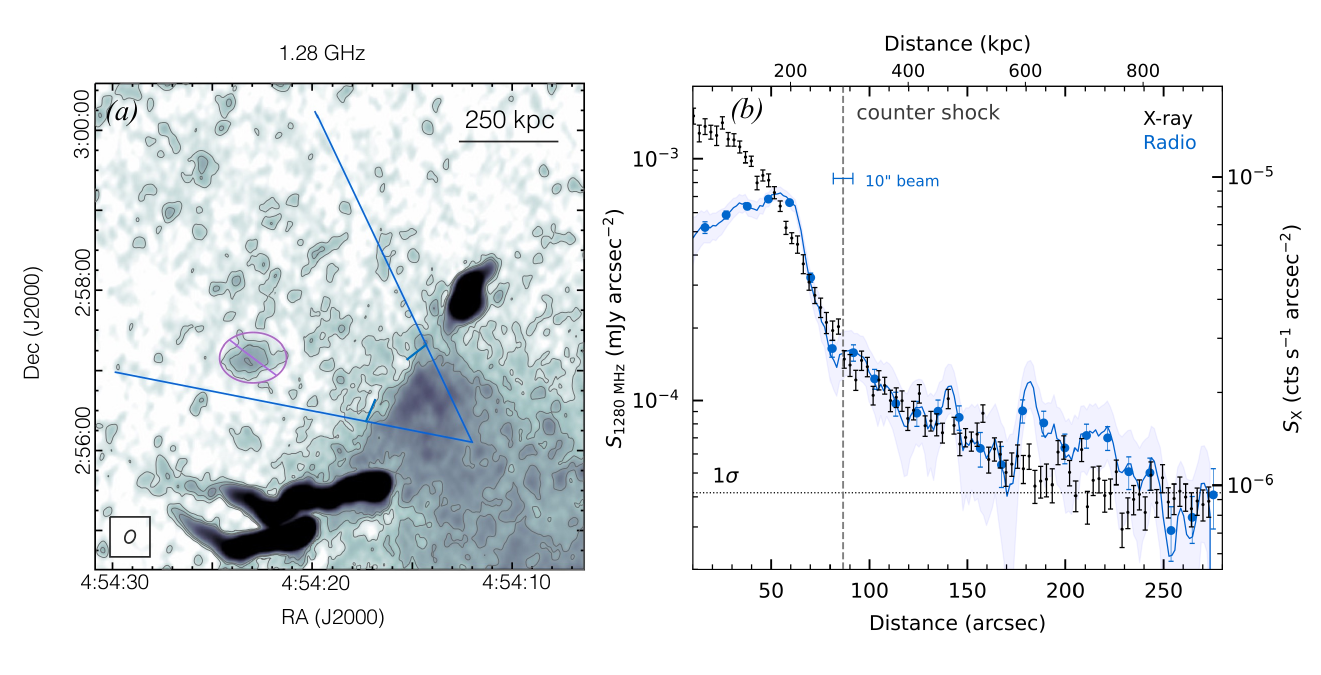}
\includegraphics[width=\hsize]{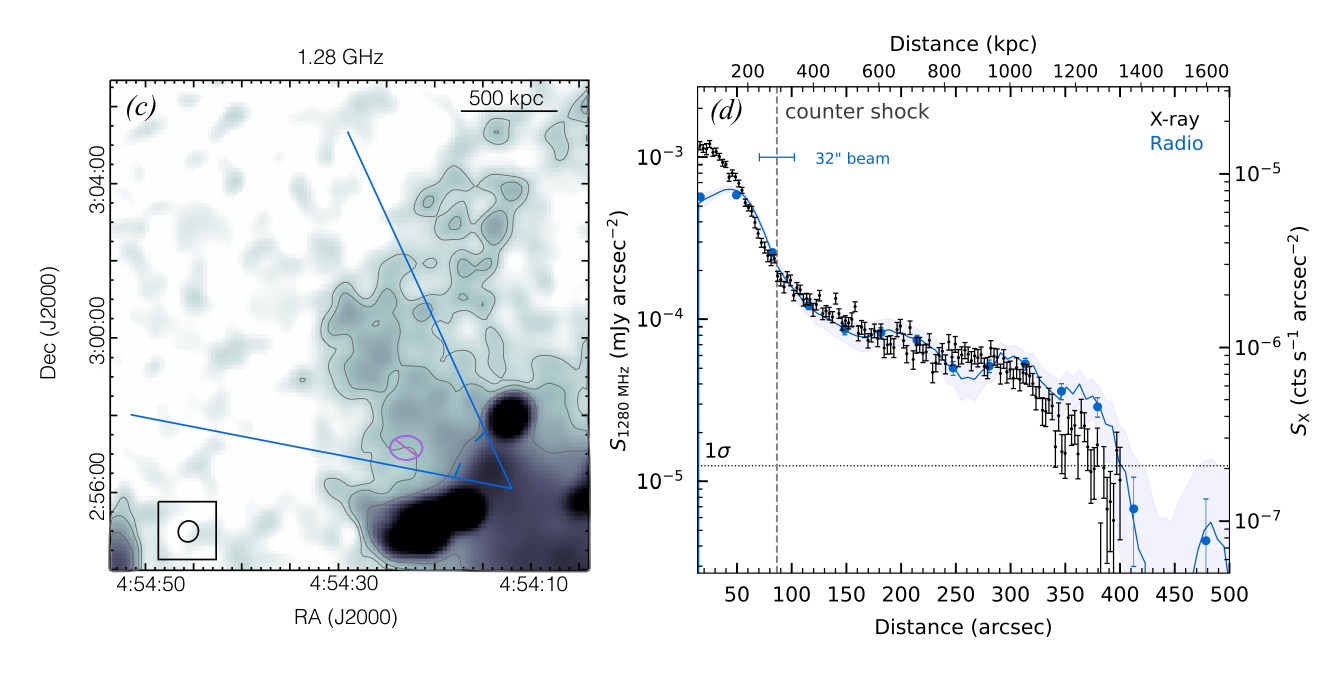}
\caption{Panels {\em (a,c)}: 1.28 GHz images of the northeast region of the radio halo at $10^{\prime\prime}\times7^{\prime\prime}$ and $33^{\prime\prime}\times31^{\prime\prime}$ resolution (boxed ellipses; $1\sigma=3.3$ $\mu$Jy beam$^{-1}$ and $14.5$ $\mu$Jy beam$^{-1}$; \#6 and 5 in Tab.~\ref{tab:images}). Contours are at 3, 4, and 5$\sigma$ in panel {\em a} and 3, 5, and 12$\sigma$ in panel {\em c}. Blue lines indicate the sector used to extract the profiles in panels {\em b} and {\em d}; short ticks show the position of the X-ray shock front. Residuals from a partially-subtracted extended source are masked (crossed ellipse). Panels {\em (b,d)}: Radio (blue) and X-ray (black) surface brightness profiles. The solid blue line and shaded region (1$\sigma$ uncertainties) are derived from radial bins of $1^{\prime\prime}.5$ in {\em b} and $6^{\prime\prime}$ in {\em d}. Blue points, with $1\sigma$ error bars, correspond to bins comparable to the beam size ($10^{\prime\prime}$ and $32^{\prime\prime}$). The horizontal dotted line indicates the radio $1\sigma$ level, and the vertical dashed line marks the counter shock position (Fig.~\ref{fig:xray}{\em a}). The X-ray drop at $\sim 400^{\prime\prime}$ is caused by the detector chip edge.} 
\label{fig:counter}
\end{figure*}

\section{Surface brightness and spectral analysis} 
\label{sec:analysis}

In this section, we examine the surface brightness and spectral behavior of the diffuse radio emission in A\,520 through radial profile analysis. 
We first focus on the two radio edges bounding the halo at the positions of the X-ray shock fronts (\S~\ref{sec:shocks}). We then analyze the large-scale diffuse emission across multiple directions from the cluster center (\S~\ref{sec:radial}) to derive a more complete picture of the extent and spectral properties of the emission components identified in \S~\ref{sec:new}.

\subsection{Radio edges and shock fronts}
\label{sec:shocks}

The radio halo is bounded by the SW and NE radio edges at the location of the X-ray bow and counter shock, respectively (Figs.~\ref{fig:hr}{\em (b)} and \ref{fig:xray}{\em (a)}). To compare these edges and the shock fronts, we extracted radial surface brightness profiles in sectors encompassing these fronts, using 
the $0.8-4$ keV {\em Chandra}\/ image (Fig.~\ref{fig:xray}{\em (a)}) after masking out  compact X-ray sources, and MeerKAT images at high ($10^{\prime\prime}\times7^{\prime\prime}$) and low ($33^{\prime\prime}\times31^{\prime\prime}$) resolution (\#6 and \#5 in Tab.~\ref{tab:images}). The sectors  are shown in Figs.~\ref{fig:bow} and \ref{fig:counter}. For the bow shock, the sector was selected to include the best-defined portion of the front in the X-ray, where the Mach number $\mathcal{M}$ is the highest (region N1+N2 in W18, with $\mathcal{M}=2.4$). 

\subsubsection{The bow shock region}
\label{sec:bow}

Figure~\ref{fig:bow} shows the radio and X-ray surface brightness profiles across the bow shock. 
 The radio emission drops sharply at the position of the X-ray shock, confirming the spatial coincidence previously established by M05, W18 and H19. The sharpness of this edge is particularly evident at the higher radio resolution (panel {\em (b)}), where the radio profile drops by more than an order of magnitude within a single beam width of the shock position.
Beyond the shock, in the pre-shock/upstream region, 
the lower-resolution radio profile (panel {\em (d)}) reveals faint emission extending $\sim300-500$ kpc ahead of the shock position. In the radio image in Fig.\ \ref{fig:hr}{\em (c,d)}, this pre-shock emission extends across the whole shock sector (i.e., is not concentrated in a few clumps), although it is barely above the noise level.

\subsubsection{The counter shock region}
\label{sec:counter}

Figure~\ref{fig:counter} shows the radio and X-ray surface brightness profiles across the NE boundary of the radio halo, opposite to the bow shock. The X-ray brightness and temperature gradients are consistent with a shock front with $\mathcal{M}=1.52$ (W16, H19), although the X-ray brightness profile does not exhibit a sharp edge similar to the SW bow shock, likely because of the less favorable projection. The high-resolution radio brightness profile (panel {\em (b)}) reveals an abrupt decrease in radio emission at the position of this shock
within approximately one beam width of the shock position, similar to the behavior seen at the bow shock. Beyond the shock, the lower-resolution profile (panel {\em (d)}) shows the radio emission of the NE tail (\S~\ref{sec:ne_tail}), detected out to $\sim1$ Mpc beyond the counter shock. It follows
the decline of the X-ray profile remarkably well (the cutoff in the X-ray profile at $\sim$400\arcsec\ is due to the {\em Chandra} detector edge). The abrupt change in surface brightness across the NE edge, from the bright emission inside the edge to the much fainter emission outside, suggests that these are two distinct components: the halo confined between the two shock fronts and the NE tail associated with the c1 and c2 mass clumps (Fig.~\ref{fig:xray}{\em (b)}, \S~\ref{sec:ne_tail}).

\begin{figure*}
\centering
\includegraphics[width=0.9\hsize]{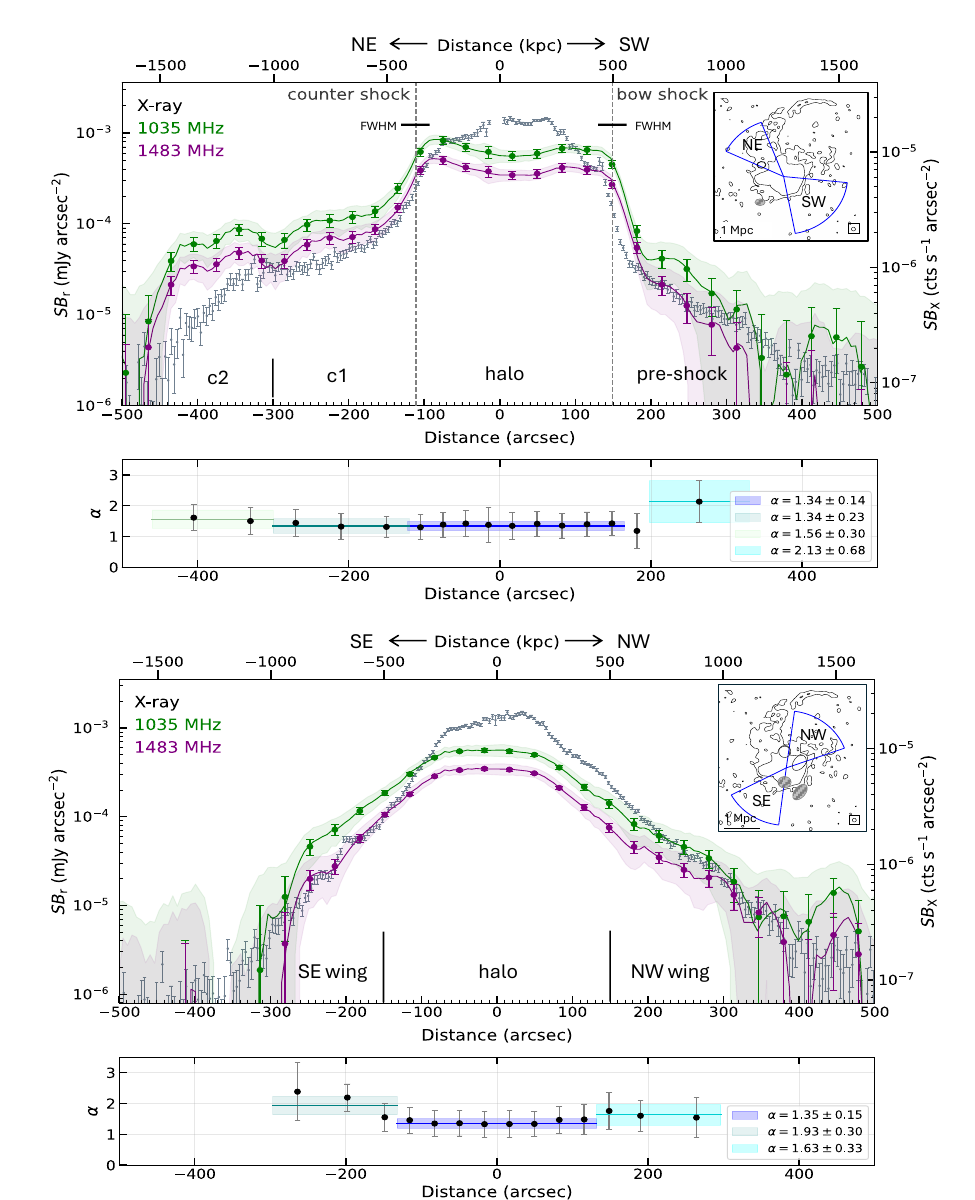}
\caption{X-ray (black) and 1035/1483 MHz (green/purple) brightness profiles computed 
using the {\em Chandra} image in Fig.~\ref{fig:xray}{\em (a)} and radio images \# 7 and 8 in Tab.~\ref{tab:images}. The sectors used for the profiles are shown in the insets, overlaid on the 1284 MHz radio contours at 2 and 12$\sigma$ (\# 5 in Tab.~\ref{tab:images}). Unsubtracted radio galaxies and X-ray point sources in the sectors were masked. The x-axis is centered on the vertex of the sectors.  Solid lines with shaded 1$\sigma$ 
regions represent $6^{\prime\prime}$ radial bins, while points with 1$\sigma$ error bars use $33^{\prime\prime}$ bins approximately matched to the radio beam ($35^{\prime\prime}\times 32^{\prime\prime}$). 1035-1483 MHz spectral index profiles are shown, with bands representing weighted averages (see legend). }
\label{fig:profile_all}
\end{figure*}

\subsection{Spectral properties of the diffuse emission}
\label{sec:radial}

To investigate the large-scale distribution of the diffuse radio emission beyond the central halo region, we extracted radial surface brightness profiles in multiple directions from the cluster center. The profiles were computed using the MeerKAT source-subtracted, low-resolution images at 1035 and 1483 MHz (\#7 and \#8 in Tab.~\ref{tab:images}) together with the {\em Chandra} 0.8–4.0 keV image (Fig.~\ref{fig:xray}{\em (a)}). The sectors used for the profile extraction are shown in the insets of Fig.~\ref{fig:profile_all} 
and were selected to probe the main large-scale features detected by MeerKAT beyond the halo, including the NE tail, the SW pre-shock emission, and the NW and SE wings. Unsubtracted radio galaxies and X-ray point sources within these sectors were masked. The resulting X-ray and radio brightness profiles are shown in Fig.~\ref{fig:profile_all}. The top pair of panels shows the NE and SW sectors, oriented along the merger axis, while the bottom pair shows the profile along the perpendicular direction.

The lower panels in each pair show the in-band spectral index profiles derived within the same sectors from the 1035 and 1483 MHz images. Colored bands represent weighted averages, whose values are reported in the legend. The errors on the spectral index include both statistical and systematic uncertainties (Sect.~\ref{sec:flux}).

Within the central $\sim1$ Mpc halo region, the spectrum is uniform on scales larger than our 100 kpc spatial bins, with an average value of $\alpha=1.34\pm0.15$, consistent with the integrated spectral index reported in Tab.~\ref{tab:diffuse}. Toward larger radii, there are hints of steepening in all directions, although the change in slope from the spectrum of the halo is statistically significant only in the direction of the SE wing, where we observe $\alpha=1.93\pm0.30$.  
For the SW pre-shock region, we estimated the spectral index by averaging four radial bins covering the range $\sim$165-550 kpc from the shock front; the bin immediately adjacent to the shock was excluded, because it falls within one beam width of the front and would be contaminated by the other side. For this newly detected pre-shock emission, we obtain the spectral index $\alpha=2.13\pm0.68$, $1\sigma$ above the index for the halo
(see also Tab.~\ref{tab:diffuse}).

\begin{figure*}
\centering
\includegraphics[width=\hsize]{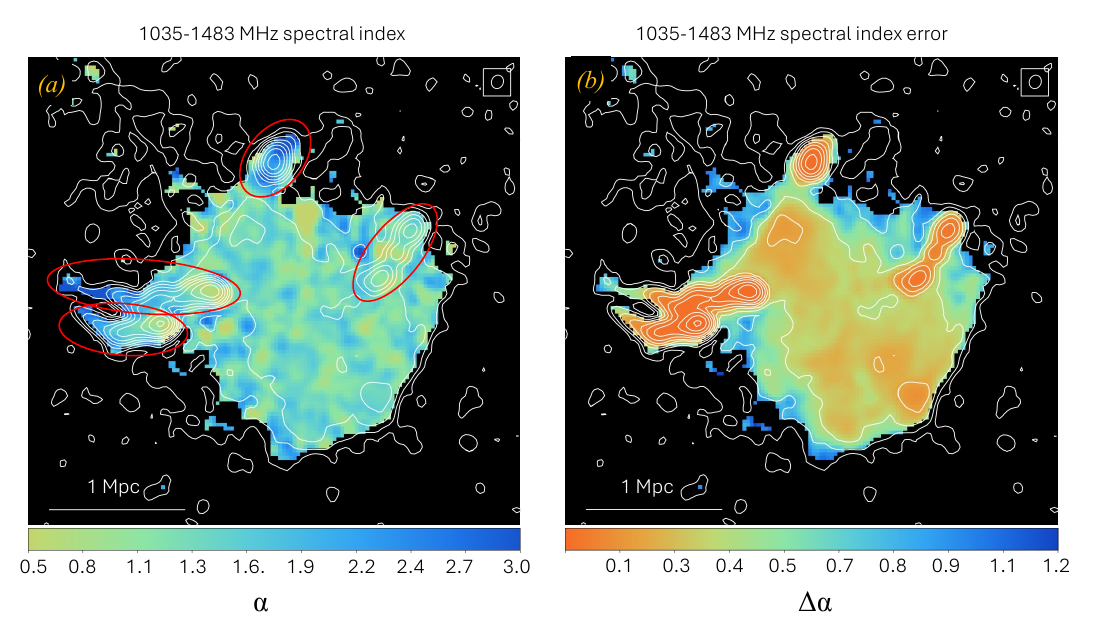}
\includegraphics[width=\hsize]{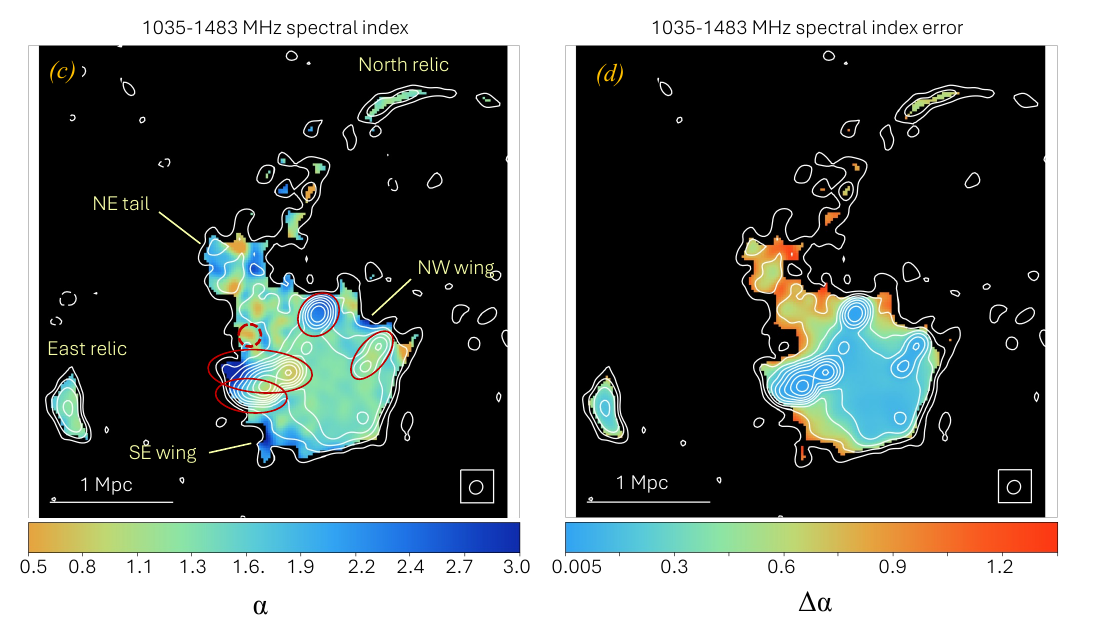}
\caption{Panels {\em (a, c)}: color-scale images of the in--band (1035-1483 MHz) spectral index distribution  $14^{\prime\prime}.7\times13^{\prime\prime}.1$ and $34^{\prime\prime}.6\times32^{\prime\prime}.1$ resolution, respectively (white, boxed ellipses) and corresponding uncertainty maps (panels {\em (b,d)}). Contours from the 1483 MHz source-subtracted image at matching resolution (\#10 and \#8 in Tab.~\ref{tab:images}) are overlaid on all color maps. Contours start at $3\sigma$ and increase by a factor of 2. Red ellipses mark four extended radio galaxies that were not subtracted, while the dashed circle indicates the location of residual faint emission from a partially subtracted extended source.}
\label{fig:spix}
\end{figure*}

\subsection{Spectral index images}
\label{sec:spix}

Figure \ref{fig:spix} shows the in--band (1035–-1483 MHz) spectral index distribution of the diffuse emission in A\,520 at two angular resolutions (panels {\em (a)} and {\em (c)}), together with the corresponding uncertainty maps (panels {\em (b)} and {\em (d)}),
in which the main contribution comes from the proximity of the frequencies (systematic uncertainties are not included). The spectral index was computed pixel-by-pixel from the images at 1035 and 1483 MHz (\#9 and \#10 in Tab.~\ref{tab:images} at high resolution, \#7 and \#8 at low resolution), which were produced with matched beams and identical shortest baseline in the $uv$ plane ($0.1$ k$\lambda$). Only pixels with surface brightness above $3\sigma$ in both images were included.

At both resolutions, the spectral index distribution across the radio halo is remarkably uniform, with values around  $\alpha \sim 1.3$ throughout most of the halo, consistent with the integrated spectral index in Tab.~\ref{tab:diffuse} and the sector profile averages in Fig.~\ref{fig:profile_all}. The apparent small-scale ($\sim 15^{\prime\prime}-20^{\prime\prime}$) fluctuations in spectral index seen at high resolution across the halo region are not statistically significant. This was determined by measuring spectral index and uncertainty values over a grid of beam-independent $20^{\prime\prime} \times 20^{\prime\prime}$ cells covering the halo ($\sim 100$ cells in total). The resulting distribution has a mean of $\langle \alpha \rangle = 1.32$ with a standard deviation of $\sigma_\alpha = 0.23$, while the mean per-cell uncertainty is $\langle \Delta\alpha \rangle = 0.35 \pm 0.12$. Therefore, the fluctuations in the map are consistent with noise and the halo spectrum being intrinsically uniform on scales of $20^{\prime\prime}$ ($\sim$ 65 kpc). H19 reached a similar conclusion from a wider frequency baseline (145 MHz--1.5 GHz) spectral index map at comparable resolution (see also V14). In addition, as a consistency check on the reliability of our spectral index maps, we note that the radio galaxies visible in Fig.~\ref{fig:spix}{\em (a)} (red ellipses) show the expected spectral index trend, with flat spectra ($\alpha \sim 0.5$) in their cores and progressively steeper spectra along their tails and lobes, consistent with synchrotron aging of the relativistic electrons as they travel away from the site of injection. The northernmost radio galaxy is characterized by a very steep spectrum ($\alpha >2$). This source was identified by V14 as a candidate dying radio galaxy (their source C).

In the low-resolution map (Fig.~\ref{fig:spix}{\em (c)}), a radial steepening is visible in both the NW/SE directions, tracing the wings, consistent with the profile analysis in Fig.~\ref{fig:profile_all}. The NE tail shows a relatively patchy spectral index distribution with average values of $\alpha \sim 1.3-1.4$ near the halo and slightly steeper moving outward. 

The low-resolution spectral index image also covers the two relics detected at large cluster radii. Both structures appear in the map with spectral index values of $\alpha \sim 1.3–1.5$, consistent with the integrated values reported in Tab.~\ref{tab:diffuse}. Due to the uncertainties, no reliable spectral trend can be identified across either relic with the current data. Finally, the bridge connecting the NE tail to the North relic is so faint that it does not allow a meaningful pixel-by-pixel spectral analysis.

\begin{figure*}
\centering
\includegraphics[width=\hsize]{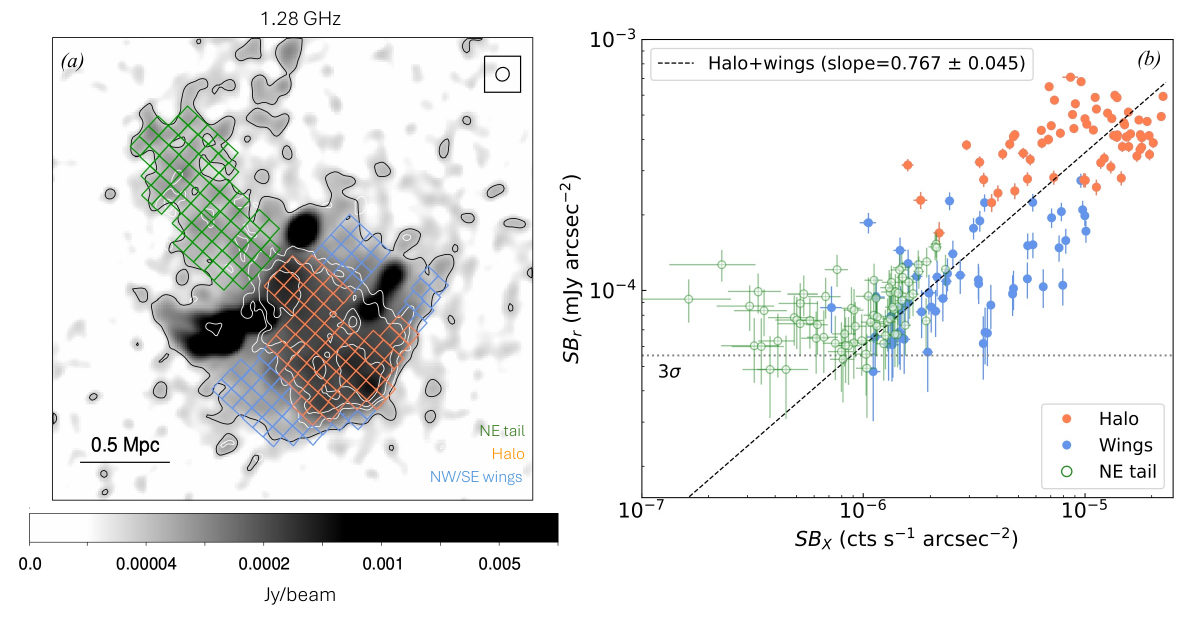}
\caption{{\em (a)} Grid used for the radio/X-ray surface brightness measurements shown in panel {\em (b)}. The grid consists of $23^{\prime\prime}\times23^{\prime\prime}$ cells and is overlaid on the MeerKAT 1.28 GHz image with a resolution of $23^{\prime\prime}.3 \times21^{\prime\prime}.7$ (boxed ellipse; \#3 in Tab.~\ref{tab:images}; $1\sigma = 10.5$ $\mu$Jy beam$^{-1}$). Orange, blue, and green represent the radio halo emission, the NW/SE wings, and the NE tail, respectively. The black contour is drawn at $+3\sigma$. White contours show the W16 source-subtracted VLA 1.4 GHz image ($+3\sigma$, increasing by factors of 2; $22^{\prime\prime}$ beam, $1\sigma = 22$ $\mu$Jy beam$^{-1}$). {\em (b)} Radio ($SB_r$)–X-ray ($SB_X$) surface brightness relation derived from the 0.8–4.0 keV {\em Chandra} image (with point sources masked; Fig.~\ref{fig:xray}{\em (a)}) and the MeerKAT 1.28 GHz image shown in panel {\em (a)}. The black dashed line indicates the best-fit relation including both the halo and the wings. The horizontal dotted line indicates the radio $3\sigma$ level.}
\label{fig:corr}
\end{figure*}

\subsection{Radio-X-ray surface brightness correlation}
\label{sec:corr}

The spatial point-to-point correlation between radio and X-ray surface brightness is a frequently-used diagnostic tool to investigate the physical connection between the relativistic and thermal components of the ICM in diffuse radio sources \citep{2001A&A...369..441G}. 
In radio halos, a tight, positive correlation between these two quantities is typically observed, which is interpreted as evidence that the distribution of relativistic electrons and/or magnetic fields is linked to the thermal gas distribution \citep[e.g.,][]{2014IJMPD..2330007B}. This analysis was previously applied to the radio halo in A\,520 by H19, who found only a weak trend between the radio and X-ray surface brightness of the halo.
The new MeerKAT data, with their substantially improved sensitivity and $uv$-coverage, allow us to revisit this analysis and extend it to the new, fainter emission components.

In Fig.~\ref{fig:corr}, we present the spatial correlation between the radio ($SB_r$) and X-ray ($SB_X$) surface brightness of the entire diffuse emission in A\,520, computed using the MeerKAT 1.28 GHz image at $23^{\prime\prime}\times 22^{\prime\prime}$ resolution (\#3 in Tab.~\ref{tab:diffuse}) and the {\em Chandra} 0.8–4.0 keV image (Fig.~\ref{fig:xray}{\em (a)}). 
We used a grid of $23^{\prime\prime}\times23^{\prime\prime}$ cells, shown in Fig.~\ref{fig:corr} with different colors to distinguish the three main components. 
 
When considering the radio halo alone (orange points), we find only marginally significant correlation between the X-ray and radio brightness, with a best-fit slope of $b=0.2$ (defined as $SB_r\propto SB_X^{\,b}$) and Spearman rank coefficient of $\rho = 0.30$ ($p=0.011$),%
\footnote{$\rho$ measures the correlation strength (ranging from $+1$ to $-1$) and $p$\/ is the probability of obtaining the observed correlation by chance under the null hypothesis of no correlation.}%
consistently with H19. However, when the halo is extended to include the fainter NW and SE wings (blue points), the correlation becomes more apparent, with a best-fit slope of $b=0.77\pm 0.05$, a Spearman coefficient of $\rho = 0.69$ and  $p=1.35\times10^{-18}$ (a fit to the wings alone gives $b=0.5\pm0.1$, $\rho = 0.62$, $p=3.6\times10^{-7}$). This supports the interpretation of the wings as an integral part of the halo system. The weak correlation previously found for A\,520 was mostly a consequence of the limited sensitivity of earlier radio observations, which missed the low surface brightness extension, in the presence of large intrinsic scatter within the halo.

The radio brightness of the NE tail (green points) is systematically above the best-fit relation defined by the halo and wings. A fit to the NE tail data points alone gives a slope of $\sim 0.2$ and a moderate Spearman correlation of $\rho = 0.56$ ($p=3.6\times10^{-7}$).
The higher radio brightness relative to the halo+wing best-fit relation suggests that the NE tail represents a physically distinct emission component, with a higher ratio of non-thermal to thermal energy density relative to the halo system. We will discuss it in \S\ref{sec:disc_dark}.

\section{Discussion}
\label{sec:discussion}

The new MeerKAT images of A\,520 uncover a remarkably complex system of diffuse radio structures, spanning scales from the cluster core out to the virial radius. In the following, we interpret these results in the context of the merger and discuss their physical implications.

\subsection{A radio halo confined by shocks}
\label{disc:halos}

As described in \S~\ref{sec:halo} and \S~\ref{sec:radial}, the radio halo in A\,520 is bracketed by two sharp radio edges coinciding with the X-ray bow and counter shocks, confining it to a $\sim$1 Mpc region along the main NE--SW merger axis. 
In the transverse direction, however, it is not confined. The newly-revealed NW and SE wings (\S~\ref{sec:extensions}) extend the halo emission smoothly by at least $\sim500$ kpc on both sides, bringing the total detected halo+wings source extent to $\sim$2 Mpc. The spectral index images and radial profiles (\S~\ref{sec:radial} and \S~\ref{sec:spix}) show that the in-band spectrum of the halo is remarkably uniform in its central region ($\alpha \sim 1.3$) and steepens toward the wings, where it reaches $\alpha \sim 1.6-2$, though with large uncertainties. 

Similar faint extensions beyond the previously-known halo boundary have been recently reported in other massive merging clusters, thanks to the high sensitivity 
of LOFAR and MeerKAT \citep[e.g.,][]{2022Natur.609..911C, 2025ApJ...992...88R}. The detection of these outermost halo regions, in some cases extending to radial distances of $\sim$2--3 Mpc with spectral steepening from the cluster core to the outskirts, is consistent with turbulent reacceleration operating over much larger volumes than previously thought \citep{2024A&A...690A..67B}. 

The radio--X-ray surface brightness correlation with slope of $\sim0.8$ found for the halo+wings source in A\,520 (\S~\ref{sec:corr}) is in line with the relations typically measured in giant radio halos \cite[e.g.,][]{2001A&A...369..441G,2005A&A...440..867G,2018ApJ...852...65R,2024A&A...686A...5B,2024ApJ...962...40S} and interpreted as a natural expectation of turbulent reacceleration models. Such sub-linear scaling implies that the non-thermal component declines more slowly with radius than the thermal ICM. Purely hadronic models, in which radio-emitting electrons are produced via CR proton interactions with the thermal ICM, predict a steeper (linear-to-super-linear) radio–X-ray correlation that more directly traces the thermal gas density distribution \citep{2001A&A...369..441G}, and are therefore disfavored by our results.

Based on the X-ray and lensing data, A\,520 appears to be a near-head-on merger of two subclusters (or more, merging along the same ``train wreck'' NE-SW axis; see Fig.\ref{fig:xray}{\em (b)}) at a stage right after the core passage (e.g., W16). According to hydrodynamic simulations \cite[e.g.,][]{kang25}, such a merger should produce two prominent shock fronts along the merger axis, propagating outward in the gas that is continuing to inflow in the wake of the subclusters. The shocked gas in the middle of the merger should also expand in the perpendicular direction, unrestrained by the above inflow, and develop an ``equatorial'' shock front encircling the merger site at a larger distance from the center. In A\,520, the two shocks on the merger axis are observed, while the more distant equatorial shock is probably beyond the current X-ray sensitivity. The gas processed and bound by this system of shock fronts should be hot and turbulent \cite[e.g.,][]{2011ApJ...726...17P}. The radio halo seen in the MeerKAT images, with sharp edges at the two shocks with some edge brightening (\S\ref{sec:halo}, Fig.~\ref{fig:profile_all})
and wide, steeper-spectrum wings perpendicular to the merger axis, is consistent with this picture --- the radio emission likely follows the spatial distribution of the merger-driven turbulence and of the seed electrons available for reacceleration.

Finally, the MeerKAT high-resolution images (e.g., Fig.~\ref{fig:hr}{\em (b)}) reveal an irregular surface brightness distribution within the halo, with 
a prominent excess toward southwest associated with the remnant cool core (W16). The latter may be related to a disrupted minihalo that inhabited the cool core prior to the merger, as suggested by W18. However, the high-resolution spectral index image (Fig.~\ref{fig:spix}{\em (a)}) shows no indication for a spectral signature distinct from the surrounding halo emission, suggesting that the aged electron population in this remnant minihalo may have been reaccelerated by the same turbulence powering the radio halo. A detailed spatial and polarimetric analysis of these fine-scale features in connection with the X-ray gas structures in the core will be presented in a forthcoming paper.

\subsection{Fossil electrons upstream of bow shock}
\label{sec:disc_preshock}

The coincidence of the sharp SW radio halo edge with the X-ray bow shock in A\,520 was first noted by M05, who proposed that the radio edge could result from either reacceleration or adiabatic compression of pre-existing relativistic electrons in the pre-shock region by the shock front. In the compression scenario, both the relativistic electrons and the magnetic field, frozen into the thermal gas, are compressed by the shock passage, boosting the radio emissivity. Such fossil electrons in the pre-shock region should produce radio emission at the brightness level that can be predicted from the observed post-shock brightness and the X-ray derived compression factor. W18 tested this prediction using a VLA dataset at 1.4 GHz. In the VLA images, the radio emission drops sharply at the X-ray shock position, with no emission detected in the pre-shock region. The VLA data should have been sensitive enough to constrain this model; however, the pre-shock region was affected by a negative interferometric artifact, likely caused by the $uv$-coverage gaps, which limited the confidence of the exclusion of the compression model to only $\sim$ 2$\sigma$.

Our new MeerKAT observation at 1.28~GHz offers a significant improvement in both sensitivity and $uv$-coverage over the previous data. The low-resolution 
radio images and radial profiles detect pre-shock emission out to $\sim$ 300--500 kpc ahead of the bow shock. Its average spectral index is $\alpha=2.1\pm0.7$, steeper than the post-shock halo region ($\alpha\sim 1.3$), as expected for fossil electrons. A quantitative test of the adiabatic compression scenario, including modeling of the pre-shock radio emission as in W18, is deferred to a future work.

\subsection{Diffuse radio emission in the dark subcluster}
\label{sec:disc_dark}

The NE radio tail, a $\sim$1.2 Mpc long structure extending along the merger axis further to the northeast from the central halo's NE edge, appears to be a distinct emission component rather than a halo extension, with an order of magnitude lower surface brightness than the halo and a high radio to X-ray surface brightness ratio (Fig.~\ref{fig:corr}, \S~\ref{sec:corr}). 

The tail contains two radio enhancements, each coincident with a mass clump in the lensing map (c1 and c2; Fig.~\ref{fig:xray}{\em (b)}) and broadly following the X-ray brightness distribution (Fig.~\ref{fig:xray}{\em (c)}). These clumps are very unusual in the X-ray, too. The outermost clump c2, could be modeled relatively well and was shown to have a very low gas mass to total mass ratio, compared to the normal cluster value --- a ``dark subcluster'' (W16). W16 also established that the gas in c2 has a high specific entropy, consistent with that of the surrounding A\,520 gas, and a higher temperature ($\sim$8 keV vs.\ $\sim$4 keV for the gas elsewhere at that radius). Its high gas temperature is far above the virial temperature for a cluster of this low mass ($\sim 2\times 10^{13}$ M$_{\odot}$, W16, \S\ref{sec:a520}). This suggests that the gas in the c2 subcluster is not a remnant of its own pre-merger gas core, which would typically have a low specific entropy, but was instead pulled in by gravity from the surrounding ICM, and adiabatically compressed, after the subcluster has emerged from the collision site as a dark matter clump stripped of gas.

This reaccretion process may also be responsible for the peculiarly bright observed diffuse radio emission. The ICM is expected to be permeated at all radii by magnetic fields and relativistic electrons of various ages, distributed across the cluster by gas motions associated with past mergers \citep{2021ApJ...914...73Z,2021A&A...653A..23V,2023A&A...669A..50V}. The gravitational pull of the dark matter clump flying through hot gas in the outskirts would create a focused perturbation and possibly significant, volume-filling turbulence within the clump and in its wake. We speculate that the turbulence in this regime, driven on small scales in a relatively uniform, high-entropy medium, might be particularly efficient for CR reacceleration. The compression of the magnetic field, advected by the compressing ICM, would also contribute to the radio enhancement. 
The spatial extent of the radio emission in the c2 region ($\sim$500 kpc) is similar to the size of the X-ray clump, and the steep radio spectral index ($\alpha\sim1.5$) is consistent with turbulent reacceleration. 
It would be interesting to confirm the presence of local turbulence with future X-ray spectrometers with high spatial resolution and sensitivity, such as NewAthena\footnote{https://www.the-athena-x-ray-observatory.eu/en/newathena-mission}.

W16 suggested that the second mass clump (c1) may be a similar object, although it is affected more strongly by projection effects, preventing a reliable thermodynamic characterization. The presence of diffuse radio emission at c1 suggests that a similar mechanism of reaccretion and compression of magnetized, relativistic plasma may also be at work there. 

We may ask whether the radio emission in the dark subcluster c2 is a small-scale radio-halo analog. The measured flux density at 1.28 GHz of $1.8\pm0.2$ mJy implies a radio power of $\sim2\times10^{23}$ W Hz$^{-1}$ at 1.4 GHz. Combined with the subcluster mass ($\sim2\times10^{13}$ M$_\odot$), this radio luminosity would place it far above the empirical radio power--mass relation found for radio halos \citep[e.g.,][]{2021A&A...647A..51C}, implying that it is extraordinarily radio-bright for its mass. On the other hand, its synchrotron {\em emissivity}\/ peak, computed following \cite{2024MNRAS.528.6470M} using an exponential fit to the radial surface brightness profile, is $\sim5\times10^{-44}$ erg s$^{-1}$ cm$^{-3}$ Hz$^{-1}$. 
This is lower than that in the center of the main halo ($\sim10^{-43}$ erg s$^{-1}$ cm$^{-3}$ Hz$^{-1}$, estimated using the same method), 
which itself lies at the faint end of the radio halo distribution, and is similar to that for the low-emissivity Coma cluster \citep[see Fig.~10 in][]{2024MNRAS.528.6470M}. However, the gas density at the center of this gas-poor clump, $n_{H}\sim1\times 10^{-3}$ cm$^{-3}$ (W16), is several times lower than the typical central density for a non-cool-core cluster hosting a giant halo. The synchrotron emissivity {\em per gas particle}\/ places clump c2 above the central halo in A\,520 (whose central density is $4\times 10^{-3}$ cm$^{-3}$, W16) and within the range of the central regions of other giant halos. Thus, the mechanism of the CR acceleration in the thermal plasma of the clump may be similar to that driving other giant radio halos, although its total radio luminosity is too high for the subcluster mass --- which may be explained by the peculiar origin of this small gas halo.

\subsection{Two relics and a bridge}
\label{sec:disc_relics}

The two newly detected relics beyond the virial radius of A\,520 share the morphology (elongated, arc-shaped), size ($0.8-1.5$ Mpc long), and spectral index ($\alpha\sim1.4-1.5$) with other peripheral relics believed to be caused by merger shocks \citep{2019SSRv..215...16V}. At $r\sim 2.4$ Mpc (the North relic) and $r\sim 2.2$ Mpc (the East relic) from the cluster center, they are among the most radially distant relics known \citep[e.g.,][]{2014MNRAS.444.3130D,2023A&A...680A..31J, 2026A&A...707A.143B, 2025ApJ...984...25R}.

In the diffusive shock acceleration (DSA) framework, the radio spectral index $\alpha$ is related to the shock Mach number $\mathcal{M}$ by $\alpha= (\mathcal{M}^2 + 1)/(\mathcal{M}^2 - 1) \equiv \alpha_\mathrm{inj} + 0.5$, where $\alpha_\mathrm{inj}$ is the injection radio spectral index
\citep[e.g.,][]{1987PhR...154....1B}. The measured integrated spectral index of the relics in A\,520 ($\alpha\sim1.4-1.5$) implies $\mathcal{M}\sim 2.2-2.5$. As commonly found for other peripheral relics, these Mach numbers may be insufficient to explain the observed radio brightness with standard DSA from the thermal pool \citep[e.g.,][]{2020A&A...634A..64B}. A more plausible scenario is reacceleration of fossil relativistic electrons by a shock \citep[e.g.,][]{2011ApJ...734...18K}. At these radii, a shock front is not possible to detect in the X-ray with the present instrumentation. Radio polarimetry from the same MeerKAT observation, which can provide direct constraints on the magnetic field orientation and strength in the relics, will be presented in a future paper.

The location of both relics is not straightforward to explain in a simple 
head-on merger scenario along the NE-SW axis. In clusters hosting double  relics, the two relics are typically found on diametrically opposite sides from the cluster center along the merger axis, tracing the outgoing shocks driven by the cluster-cluster collision \citep{2019SSRv..215...16V}. In A\,520, the relics are not symmetric with respect to the cluster center and are offset from the merger axis. One possibility is another merger: W16 suggested that A\,520 may experience a separate merger along the north-south direction, based on structure in the X-ray images, which may drive its own pair of outgoing shocks.

Another possibility is hinted at by the giant, faint bridge, an intriguing new structure that we see connecting the c2 clump and the North relic, with the relic aligned to be the tip of this structure (Fig.~\ref{fig:hr}{\em (c)}). The spectral index of the bridge ($\alpha\sim1.4$) is consistent within errors with both the end of the NE tail ($\sim1.5$) and the North relic ($\sim1.4$). We speculate that in addition to the two observed shock fronts in the cluster inner region, the NE-SW cluster merger drives a shock front at a larger distance, near the cluster virial radius, as expected for major mergers.  This giant front, propagating along the merger axis, is charted by the North relic, the bridge, the outer end of the NE tail, and the East relic (Fig.\ref{fig:hr} {\em (c,d)}). The large variation in the radio brightness between the two relics and the bridge, and the gap between c2 and the East relic, would reflect the non--uniformity of the density of the seed CR electrons that are re-illuminated by the passage of this shock, with the two relics being regions of particularly high fossil CR density --- e.g., faded, stretched radio bubbles \citep[e.g.,][]{2021ApJ...914...73Z}.

In this scenario, a future high-sensitivity X-ray imaging instrument may detect an X-ray brightness edge spanning these structures. The apparent spatial connection between the NE tail and the bridge (a 90\deg\ turn) would be a coincidence, though not entirely, since the subcluster c2 should be a part of the merger that is driving the shock front.
The front delineated by these structures is not a perfect arc concentric with the cluster center, which is as expected because of the continuing inflow of gas along the NE-SW filament (see, e.g., the discussion of the Bullet cluster  Mach cone opening in simulations by \citealt{springel07}).

An alternative scenario for the bridge is that it is the central ridge of a large-scale structure filament, crossing the main NE-SW filament that is the main chain of subclusters. Its fossil CR electrons are re-illuminated by shocks and turbulence filling the volume of the merger. Such a configuration would not be unprecedented: radio bridges connecting halos to peripheral relics have been observed in a small number of systems \citep[e.g.,][]{2021ApJ...907...32B,2022A&A...660A..81V,2025A&A...695L..16S} and are interpreted as tracers of large-scale magnetic fields and particle distributions in cluster outskirts. In this scenario, the ICM density should also be enhanced in the bridge, which may be detectable by a future high-sensitivity X-ray imager. A future wide-field weak lensing map (e.g., from Roman Space Telescope) may also uncover the dark matter concentration along the bridge.

\section{Summary and conclusions}

We have presented deep MeerKAT radio observations of the 
merging galaxy cluster A\,520 at 1.28 GHz, reaching an rms 
noise level of $\sim$3 $\mu$Jy beam$^{-1}$ at $9^{\prime\prime}\times 6^{\prime\prime}$ resolution. Such high sensitivity, coupled with the superb $uv$-coverage of MeerKAT, have allowed us to image the diffuse radio halo permeating the cluster core with unprecedented detail and 
to uncover a wealth of new, remarkable diffuse structures. These new features include faint NW and SE extensions of the radio halo, a $\sim$1.2 Mpc tail to the northeast (NE tail), two radio relics (North and East relic), a faint bridge connecting the NE tail to the North relic, and very low surface brightness emission ahead of the X-ray bow shock. These structures extend from the cluster core out to and beyond the virial radius, revisiting current assumptions about the merger geometry. Our main results are summarized below.

\begin{itemize} 

\item[--]{\em The radio halo and NW-SE wings.} The central Mpc-sized radio halo is sharply bounded by two radio edges tracing the bow and counter X-ray shocks, perpendicular to the NE-SW merger axis. The halo has an irregular, asymmetric surface brightness distribution, with enhanced emission near the shocks and a prominent excess toward the southwest, associated with the remnant cool core. The spectral index is uniform ($\alpha \sim$1.3) across the halo's entire extent, with no significant substructures or trends visible in the spectral index images and profiles. Faint diffuse emission, extending $\sim$500 kpc to the northwest and southeast of the halo, is detected for the first time, with steep spectral indices of $\sim$1.6--2. These NW and SE wings show a smooth morphological continuity with the inner halo, with no sharp boundaries, suggesting that they are lower surface brightness and steeper-spectrum extensions of the halo to larger cluster radii. Including the wings, the total extent of the halo reaches $\sim$2 Mpc in the NW--SE direction. 

The point-to-point radio–X-ray surface brightness correlation for the halo alone is weak and flat (slope $\sim0.2$), 
consistent with the earlier result of H19. When the NW and SE wings are included, the correlation becomes significantly stronger (slope $\sim0.8$, Spearman coefficient $\rho=0.7$, chance probability $p= 1.35\times10^{-18}$). This further indicates that the wings are an integral part of the halo system and that the weak correlation previously reported for the halo was caused by the missing low surface brightness emission at larger cluster radii. The sub-linear slope found for the halo+wing system favors turbulent reacceleration over purely hadronic models. 

\item[--]{\em SW pre-shock emission}. Very faint diffuse emission is detected ahead of the X-ray bow shock, out to $\sim300-500$ kpc into the undisturbed preshock gas and with a steep spectral index of $\alpha = 2.1 \pm 0.6$. This detection may indicate the presence of fossil relativistic electrons in the upstream region, expected in the adiabatic compression / reacceleration model considered by M05 and W18 for the origin of the radio emission at the bow shock. A quantitative test of this scenario is deferred to a forthcoming paper.

\item[--] {\em The NE tail and the dark subcluster}. A $\sim 1.2$ Mpc-sized tail of diffuse emission is discovered to the northeast of the halo, separated from it by the NE shock edge and likely representing a distinct emission component. This is further supported by the point-to-point radio/X-ray surface brightness analysis, where the NE tail is partially offset above the halo+wings relation, with a mild, shallower correlation of its own. The NE tail is aligned with the main merger axis and is spatially coincident with two mass clumps, c1 and c2, identified in weak lensing and X-ray images. The radio tail shows two distinct radio enhancements, each associated with one of these clumps, with $\alpha\sim1.3$ at c1 and $\sim1.5$ at c2. The outermost clump (c2) is the so-called {\em dark subcluster}, a relatively low-mass ($\sim2\times10^{13}$ M$_{\odot}$), gas-poor post-merger remnant, which was stripped of its gas during the crossing of the A\,520 core and is now re-accreting the surrounding cluster gas after emerging from the collision site. We propose that, as the dark subcluster re-accretes the ambient ICM along with the non-thermal plasma mixed with it, adiabatic compression and turbulent reacceleration within the subcluster produces the observed diffuse radio emission. A similar mechanism may operate at the clump c1. The radio emissivity per gas particle in the dark subcluster is similar to that in giant radio halos, consistent with the same reacceleration mechanism responsible for the radio emission. The total radio luminosity of the clump, too high for its low mass, sets it far apart from other radio halos.

\item[--]{\em The radio relics}. Two relics are detected for the first time in A\,520, both located beyond the cluster virial radius. The North relic is at $\sim$2.4 Mpc from the X-ray centroid, with a thin arc-shaped morphology spanning $\sim$1.5 Mpc and with $\alpha\sim1.4$. The East relic is at a distance of $\sim2.2$ Mpc, with a largest size of $\sim0.8$ Mpc and $\alpha\sim1.5$. These are among the most radially distant peripheral relics known. 
Their locations are difficult to reconcile with a simple head-on merger along the NE–SW axis, and may instead reflect a more complex merger geometry involving at least a secondary collision along the north–south direction, as suggested by W16. An alternative possibility is given below.

\item[--]{\em The giant radio bridge}. The NE tail ends abruptly at the dark subcluster and connects, along a direction perpendicular to the main merger axis, to the North relic via a faint, $\sim2$ Mpc long radio bridge. The NE tail, bridge, and North relic may form a physically connected structure tracing the underlying filament of the large scale structure, crossing the main NE-SW filament. Future sensitive weak lensing and X-ray imaging observations may detect this filament. An alternative possibility is another giant shock front, driven by the same NE-SW cluster merger and propagating along the merger axis, presently located near the virial radius. The North relic, the bridge, the end of the NE tail, and the East relic would delineate this shock front, with the variations in radio brightness reflecting the very non--uniform density of the fossil CR electrons that are reaccelerated by the passage of this giant shock. Future wide-field X-ray imaging may detect this shock front as an X-ray brightness edge.

\end{itemize}

These results reveal that A\,520 is a significantly more extended and morphologically complex system than previously known, with diffuse radio emission spanning scales from the cluster core out to $r_{\rm 200}$. The detection of these structures has been made possible by the exceptional sensitivity of MeerKAT to extended low surface brightness emission. The upcoming Square Kilometre Array (SKA), with its order-of-magnitude improvement in sensitivity and $uv$-coverage over current facilities, promises to reveal an even richer picture of diffuse radio emission in clusters
\citep{2026arXiv260620366B}. 

The results presented here open several possibilities for future investigation. The MeerKAT observations of A\,520 were carried out in full polarization mode, and an analysis of the polarimetric properties of the relics and halo edges, which would provide evidence for ordered magnetic fields at shock locations, is planned as part of a forthcoming paper. A quantitative test of the adiabatic compression scenario for the SW preshock emission, including the detailed modeling of the expected pre-shock radio brightness as a function of the shock parameters, is also deferred to a forthcoming paper, along with a detailed analysis of the radio and X-ray properties of the cool core remnant and its possible association with a disrupted minihalo. Observations with a wider frequency coverage, combining, for example, uGMRT observations with deep MeerKAT's UHF band (800 MHz) and S--band (3 GHz) data, would substantially improve the spectral index constraints of all emission components, which are currently limited by the narrow in-band frequency leverage of the MeerKAT 1.28 GHz data alone.  Finally, deeper X-ray observations would be invaluable for detecting surface brightness discontinuities at the locations of the peripheral relics and for characterizing the thermal gas properties in the NE tail and bridge region, but are beyond the capabilities of the current \textit{Chandra} and \textit{XMM-Newton} data. ESA's \textit{NewAthena}, currently in its study phase and with possible launch in the mid-2030s, will be able to provide substantially improved collecting area and spectral resolution that may make such measurements feasible for the first time.

\begin{acknowledgments}
We thank the referee for their helpful comments. Basic research in radio astronomy at the Naval Research Laboratory is supported by 6.1 Base funding. The MeerKAT telescope is operated by the South 
African Radio Astronomy Observatory, 
which is a facility of the National Research Foundation, an agency of the Department of Science and Innovation. The authors acknowledge the contribution of those who designed and built the MeerKAT instrument. The financial assistance of the South African Radio Astronomy Observatory (SARAO) towards this research is hereby acknowledged (www.sarao.ac.za). The National Radio Astronomy Observatory and the Green Bank Observatory are facilities of the National Science Foundation, operated under a cooperative agreement by Associated Universities, Inc.
We would like to thank Dr. Wendy Peters (U.S. Naval Research Laboratory) for sharing her expertise on peeling and imaging the radio data using OBIT.
\end{acknowledgments}

\facilities{MeerKAT, CXO, XMM}

\software{OBIT \citep{2008PASP..120..439C}, CASA \citep{2022PASP..134k4501C}, WSClean \citep{2014MNRAS.444..606O}, AIPS \citep{2003ASSL..285..109G}, DS9 \citep{2003ASPC..295..489J}.}

{}

\end{document}